\documentclass[11pt]{article}
\usepackage{amsmath,amssymb,graphicx,multirow,color,cite,bbold}

\def\be{\begin{equation}}
\def\ee{\end{equation}}
\def\bea{\begin{eqnarray}}
\def\eea{\end{eqnarray}}
\def\UMNS{U^\text{MNS}}

\def\SMS{\frac{\Delta m_{\tilde\ell}}{m_{\tilde\ell}}(\tilde e_L, \tilde\mu_L)}
\def\SMSt{\Delta m_{\tilde\ell}(\text{\small $\tilde e_L, \tilde\mu_L$}) / {m_{\tilde\ell}}}

\def\One{\mathbb{1}}

\newcommand{\GUT}{\text{GUT}}
\newcommand{\TeV}{\text{TeV}}
\newcommand{\GeV}{\text{GeV}}
\newcommand{\eV}{\text{eV}}
\newcommand{\rDM}{\Omega_\text{DM} h^2}
\newcommand{\figref}[1]{Fig.~\ref{#1}}

\newcommand{\secref}[1]{Sec.~\ref{#1}}
\renewcommand{\eqref}[1]{Eq.\,(\ref{#1})}

\begin{document}

\vspace*{-2cm}
\begin{flushright}
CFTP 13-022\\
PCCF RI 13-06\\

\vspace*{2mm}
\today
\end{flushright}
\begin{center}
\vspace*{10mm}
{\Large\bf Slepton mass splittings and cLFV in the SUSY seesaw in the light of recent
 experimental results} \\
\vspace{1cm}
{\bf A. J. R. Figueiredo$^{a,b}$ and 
A. M. Teixeira$^{b}$
}

 \vspace*{.5cm} 
$^{a}$ Centro de F\'{\i}sica Te\'orica de Part\'{\i}culas, 
Instituto Superior T\'ecnico, \\ Av. Rovisco Pais 1, 
1049-001 Lisboa, Portugal

\vspace*{.2cm} 
$^{b}$ Laboratoire de Physique Corpusculaire, CNRS/IN2P3 -- UMR 6533,\\ 
Campus des C\'ezeaux, 24 Av. des Landais, F-63177 Aubi\`ere Cedex, France

\end{center}

\vspace*{5mm}
\begin{abstract}

Following recent experimental developments, in this study we 
re-evaluate if the interplay of high- and low-energy lepton flavour
violating observables remains a viable probe to test
the high-scale type-I supersymmetric seesaw.
Our analysis shows that
fully constrained supersymmetric scenarios no longer allow to explore
this interplay, since recent LHC data precludes the possibility of
having sizeable slepton mass differences for a 
slepton spectrum sufficiently light to be produced, and in association to
BR($\mu\to e\gamma$) within experimental reach. 
However, relaxing the strict universality of supersymmetric
soft-breaking terms and fully exploring heavy neutrino dynamics, 
still allows to have slepton
mass splittings $\mathcal{O}$(few \%), for slepton masses accessible
at the LHC, with associated $\mu\to e\gamma$ rates
within future sensitivity. For these scenarios, we illustrate how 
the correlation between high- and low-energy  lepton flavour
violating observables allows to probe
the high-scale supersymmetric seesaw.

\end{abstract}

\section{Introduction}

Supersymmetric (SUSY) seesaw realisations offer an appealing framework to
address several of the observational and theoretical shortcomings of
the Standard Model (SM). Even if realised at a very
high-scale (close to the grand unification scale, $M_\GUT$),
prior to their decoupling, the
new right-handed neutrino superfields induce corrections into the 
SUSY soft-breaking slepton terms. 
Since neutrino oscillations do not conserve lepton flavour, 
these corrections are lepton flavour violating (LFV), and can induce 
SUSY contributions to slepton mediated charged LFV
(cLFV) observables~\cite{Borzumati:1986qx}. 

Compared to its non-SUSY version~\cite{seesaw:I}, and 
in addition to accounting for
neutrino masses and mixings, the high-scale type-I SUSY seesaw
opens the door to a large number of cLFV observables at/below the TeV
scale, that can be searched for in low-energy, high intensity
facilities or in high-energy colliders as the LHC or a future Linear
Collider (LC). Among the former one has flavour violating radiative
and three body lepton decays, as well as 
muon-electron conversion in nuclei~\cite{Hisano:1995cp, Hisano:1995nq, Hisano:1998fj,
  Buchmuller:1999gd, Kuno:1999jp, Ellis:1999uq, Hisano:2001qz,Casas:2001sr,
  Lavignac:2001vp, 
  Bi:2001tb, Ellis:2002fe,Deppisch:2002vz, Fukuyama:2003hn,
  Brignole:2004ah,  Masiero:2004js, Fukuyama:2005bh,
  Petcov:2005jh,Arganda:2005ji, Deppisch:2005rv, Yaguna:2005qn,
  Calibbi:2006nq,Antusch:2006vw,Arganda:2007jw, Arganda:2008jj}; 
the latter are associated to the potential reconstruction of 
SUSY decay chains involving slepton intermediate states, and include
various observables, such as for example 
flavoured slepton mass differences and 
direct flavour violating gaugino
decays~\cite{Arkanihamed:1996au,Hinchliffe:2000np,Carvalho:2002jg,Buckley:2006nv,Hirsch:2008dy,Carquin:2008gv,Esteves:2009vg,Buras:2009sg,Abada:2010kj,Abada:2011mg,Calibbi:2011dn,Galon:2011wh,Arbelaez:2011bb,Abada:2012re,Cannoni:2013gq}.   

However, in the absence of SUSY discovery (and reconstruction of its
fundamental Lagrangian), the contributions to the different 
cLFV observables allow for a wide range of predictions, as the
observables are in general dependent on powers of the average SUSY
scale and of the seesaw scale. While the first might be possibly known in the
near future, the second cannot be directly probed, which renders these
scenarios hard to test. However, 
when embedded into flavour blind SUSY breaking models, 
the type-I seesaw is the unique
source of flavour violation in the lepton sector, implying that all 
lepton flavour violating observables will be correlated. The study of the 
synergy between different low-energy observables and/or high-energy
ones proves to be a powerful tool to probe the high-scale type-I SUSY
seesaw (see, for 
example,~\cite{Buckley:2006nv,Carquin:2008gv,Esteves:2009vg,Abada:2010kj,
Calibbi:2011dn,Abada:2012re}). 

Since the first related analyses, important experimental developments have
occurred, in a number of fronts. 
Firstly, $\theta_{13}$ has been
measured~\cite{Abe:2011sj,Abe:2011fz,Ahn:2012nd,An:2012bu}, its value
being sizeable. Regarding high-energy experiments, 
LHC negative searches on SUSY particles suggest a
considerably heavier SUSY
spectrum~\cite{ATLAS-CONF-2013-047,LHC.3rdgeneration.squarks,CMS-PAS-SUS-12-022,ATLAS-CONF-2013-028,ATLAS-CONF-2013-049,Chatrchyan:2013oca},
which puts increasingly stronger 
bounds on the parameter space of constrained SUSY models. Accommodating 
the measured mass of the recently discovered 
SM-like Higgs boson~\cite{LHC.higgs} renders the latter bounds even
more severe. Finally, 
the MEG experiment has significantly improved the upper bounds on
BR($\mu \to e \gamma$)~\cite{Adam:2013mnn}.
In view of the latter developments, it is important to re-evaluate the
prospects of probing the type-I SUSY seesaw via the synergy between 
slepton mass differences (if measured at the LHC) and low-energy cLFV
observables such as BR($\mu \to e \gamma$). 
Charged sleptons may indeed be discovered in the forthcoming $\sqrt s
= 14$ TeV LHC run or then
in the subsequent high luminosity phase, for which an integrated
luminosity  
$\sim 3000$ fb$^{-1}$ is expected~\cite{Heuer:2012gi,Barr:1512933}. 
If indeed discovered, promising windows over the lepton flavour puzzle
can be opened, with prospects for shedding light on the mechanism 
of neutrino mass generation\footnote{A recent work has revisited
  charged cLFV signatures, within a SU(5) GUT framework, in low-energy
  observables and in flavour violating neutralino
  decays~\cite{Cannoni:2013gq}. Low-energy cLFV in the framework of an SO(10) embedded 
  type-I SUSY seesaw, taking into account the constraints from $m_h$,  
  was discussed in~\cite{Calibbi:2012gr}.}.  

The aim of the present study is thus to 
discuss whether sleptons with inter-generational mass 
differences (resulting mainly from a high-scale type-I SUSY seesaw), 
compatible with current cLFV results and negative SUSY searches, 
can be seen in future LHC runs, and how such observations would in turn 
affect the information one could derive on the seesaw parameters. 
To do so, we consider the embedding of a type-I seesaw into
constrained SUSY models, in particular into the constrained minimal
supersymmetric standard model (cMSSM), extended by three generations of
right-handed neutrino superfields. We then relax some of the cMSSM
strict universality conditions for the different sectors, still
preserving flavour universality. We discuss the impact of these
scenarios on high-energy cLFV observables as slepton mass differences
(between sleptons of different families), while at low-energies we focus on 
$\mu \to e \gamma$ decays and $\mu - e$ conversion in Nuclei.

The paper is organised as follows. In Section~\ref{sec:model} 
we briefly describe the 
type-I SUSY seesaw model and its most relevant phenomenological signatures. 
In Section~\ref{sec:results} we present the analysis
and discuss the results; our conclusions are 
summarised in Section~\ref{sec:concl}.

\section{The SUSY seesaw model} \label{sec:model}

The type-I SUSY seesaw consists of the Minimal Supersymmetric Standard
Model (MSSM), extended by three generations of right-handed neutrino (chiral)
superfields $\hat N^c_i \sim \left( \nu^c, \tilde \nu^*_R \right)_i$. 
The leptonic part of the superpotential reads
\be\label{eq:superpotential:lepton}
\mathcal{W}^{\text{lepton}} \,=\,  \hat N^c \,{Y}^\nu \,\hat L\, \hat H_2 + 
\hat E^c \,{Y}^l\, \hat L \,\hat H_1 + \frac{1}{2} \hat N^c \,{M_R}\, \hat N^c\,,
\ee
where $\hat L$ and $\hat E^c$ denote the SU(2) lepton doublet and
right-handed charged lepton superfields, respectively, and $\hat H_{1,2}$ are
the two Higgs supermultiplets. Without loss of
generality, we work in a basis where both  
the charged lepton Yukawa couplings ${Y}^l$ and the 
Majorana mass matrix ${M_R}$ are diagonal. For 
completeness\footnote{Since we work in a regime $M_R \gg m_\text{soft}$
the effects of the $B_\nu$-term (assumed to be $B_\nu \sim m^2_\text{soft}$)  become 
negligible in comparison to the superpotential mass term ($\tilde \nu^*_R M^2_R \tilde \nu_R$) 
-- see for e.g.~\cite{Grossman:1997is} for a discussion --, and will not be taken 
into account in the analysis.}, the  
slepton soft breaking potential is given by
\bea\label{eq:vsoft:lepton}
\mathcal{V}^{\text{slepton}}_{\text{soft}} & = 
& \tilde \ell^*_L \,{m}^2_{\tilde L} \,\tilde \ell_L + 
\tilde \ell^*_R \,{m}^2_{\tilde E} \,\tilde \ell_R + 
\tilde \nu^*_R \,{m}^2_{\tilde \nu_R}\, \tilde \nu_R \nonumber\\
&& + \left( \tilde \ell^*_R \,{A}^l \,\tilde \ell_L\, H_1 + 
\tilde \nu^*_R \,{A}^\nu \,\tilde \nu_L \,H_2 + 
\frac{1}{2} \tilde \nu_R \,{B}_\nu \,\tilde \nu_R + \text{H.c.} \right) \, .
\eea

We consider a flavour-blind
SUSY breaking mechanism (so that the Yukawa couplings are the
only source of flavour violation), as for example the case of minimal 
supergravity mediated SUSY breaking,
assuming that the soft breaking parameters satisfy 
universality conditions at some high-energy scale, 
which we take to be the gauge coupling 
unification scale, $M_\GUT \sim 10^{16}$~GeV:
\be\label{eq:msugra:conditions}
M^\psi_{i}\,=\,M_{1/2}\,,
\quad
( m^2_{\tilde \phi} )_{ij} \,=\, \delta_{ij} m^2_0\,,
\quad
( A_{\phi} )_{ij} \,=\, A^\phi_0 \, (Y^\phi)_{ij}\,.
\ee
	
In the seesaw limit (i.e., $Y^\nu v_2 \ll M_R$), after 
electroweak (EW) symmetry breaking, the light neutrino mass matrix is
approximately given by $m_\nu \simeq -v^2_2 {Y^\nu}^T M^{-1}_R Y^\nu$, 
where $v_2$ is one of 
the vacuum expectation values of the neutral Higgs $H_i$ 
($v_{1(2)} = v \cos(\sin)\beta$, with $v = 174$~GeV).
As suggested from the seesaw expression for $m_\nu$, 
a convenient means of parameterising the neutrino 
Yukawa couplings $Y^\nu$, while at the same time allowing 
to accommodate neutrino data, is given by the Casas-Ibarra
parameterisation~\cite{Casas:2001sr}.   
At the seesaw scale one can write  
\be
Y^\nu\, =\, \frac{i}{v_2} \sqrt{M^\text{diag}_R} \, R \, 
\sqrt{m^\text{diag}_\nu} \, {U^\text{MNS}}^\dagger \,, \label{eq:CasasIbarra}
\ee
which we will use in our
numerical analysis. In the above, 
$U^\text{MNS}$ is the leptonic mixing matrix and 
$R$ is a complex orthogonal matrix, parameterised in terms of three
complex angles ($\theta_i$), 
that encodes additional mixings involving the right-handed (RH) neutrinos; 
$m^\text{diag}_\nu$ and $M^\text{diag}_R$ 
respectively denote the (diagonal) light and heavy neutrino mass
matrices. 

\subsection{Flavour violation in the slepton sector}
Due to the non-trivial flavour structure of $Y^\nu$, the running from $M_\GUT$ down to the 
seesaw scale $M_R$ will induce flavour mixing in the otherwise (approximately) 
flavour conserving slepton soft breaking
terms~\cite{Borzumati:1986qx}. 
This running is more pronounced in the ``left-handed'' soft breaking 
terms (i.e., the terms involving slepton doublets).
At leading order (leading logarithm (LLog) approximation), 
the flavour mixing induced by the renormalisation group (RG) flow reads 
\bea
&& \left( \Delta m^2_{\tilde L} \right)_{ij} \,=\, -\frac{1}{8\pi^2} \left( 
m^2_{\tilde L} + m^2_{\tilde \nu_R} + m^2_{H_2} + \left| A^\nu_0 \right|^2
\right) \left( {Y^\nu}^\dagger L Y^\nu \right)_{ij} \,, \label{eq:LLogDeltaM2} \\
&& \left( \Delta A^l \right)_{ij} \,=\, -\frac{1}{16\pi^2} 
\left( A^l_0 + 2 A^\nu_0 \right) Y^l_{ii} 
\left( {Y^\nu}^\dagger L Y^\nu \right)_{ij} \,;
\, L_{kl} \equiv \log\left(\frac{M_\GUT}{M_{R_k}}\right) \delta_{kl} \,.\label{eq:LLogDeltaA0}
\eea
As is clear from the above, the amount of flavour violation in  
the slepton sector is encoded in $\left( {Y^\nu}^\dagger L Y^\nu
\right)_{ij}$, originating
from light neutrino mixing and from possible mixings
involving the heavy neutrinos (see~\eqref{eq:CasasIbarra}).
Having a unique source of LFV is the key to all tests of the
SUSY seesaw; this becomes particularly clear in the simple
(conservative) limit in which one assumes little (or no) 
additional mixing involving the
heavy RH states (i.e., $R\sim\One$). To a good approximation, 
the intrinsic amount of cLFV is 
related to low-energy leptonic mixings as
\be
\left( {Y^\nu}^\dagger L Y^\nu \right)_{ij}
\simeq 
 \UMNS_{ik} {\UMNS_{jk}}^* \left( m_k M_{R_k} L_k \right) \,.
\ee
Considering ratios of cLFV observables with similar loop dynamics -
approximately equal to ratios of the above quantity -, firstly allows
to test the SUSY seesaw by checking whether or not its degrees of
freedom can accommodate the value of (future) measured quantities.
In turn, this may then allow 
to extract information on the heavy spectrum, $M_{R_k}$ 
(although there is still a dependence on the neutrino mass hierarchy
and $\UMNS$ phases).
On the other hand, by comparing observables with different loop dynamics 
but similar flavour structure (e.g., $\ell_i \to 3\ell_j$ and 
$\ell_i \to \ell_j\gamma$) one may test new sectors where LFV can be present.

\subsection{Slepton induced cLFV observables} \label{sec:cLFVobservables}
Slepton flavour mixing can lead to charged lepton flavour violation, manifest
in a wide array of observables, at both low-energies (rare processes
searched for at high-intensity experiments, such as MEG and BaBar) 
and high energies (at colliders, 
above the slepton production threshold). Having one unique source
of flavour violation implies that the observables should exhibit some
correlation which, as extensively discussed in the
literature (see, for
example,~\cite{Buckley:2006nv,Carquin:2008gv,Esteves:2009vg,
Abada:2010kj,Calibbi:2011dn,Abada:2012re}), 
allows to indirectly probe the high-scale seesaw
hypothesis. 

\medskip
At low-energies, virtual sleptons can mediate flavour violating lepton
transitions, such as radiative decays,
three-body decays and conversion in nuclei. As an example, 
the radiative decay $\ell_i \to \ell_j \gamma$ receives 
contributions originating from sneutrino-chargino and charged
slepton-neutralino loops (see e.g.~\cite{Raidal:2008jk}, and
references therein). Compared to the SM contributions, 
which are highly suppressed by powers of $m_\nu/M_W$, 
these new contributions can be sizeable provided 
$m_{\tilde\ell_L}$ is not {\it too heavy} and 
slepton flavour mixing is large. 
An analytical understanding of the 
dependency of BR($\ell_i \to \ell_j \gamma$) 
on the neutrino Yukawa couplings 
can be obtained using the LLog 
approximation. In the limit of very small off-diagonal 
$\Delta m^2_{\tilde L} $ entries, one has
\be                 
\frac{\text{BR($\ell_i\to\ell_j\gamma$)}}
{\text{BR($\ell_i\to\ell_j\nu_i\bar\nu_j$)}} \,\approx\,  
\frac{\alpha^3 \,\tan^2\beta}{G^2_F \,m^8_\text{SUSY}} 
\left| \frac{1}{8 \pi^2} 
\left( m^2_{\tilde L} + m^2_{\tilde \nu_R} + 
m^2_{H_2} + \left| A^\nu_0 \right|^2 \right) 
\left( {Y^\nu}^\dagger \,L \,Y^\nu \right)_{ij} \right|^2 \,. \label{eq:BRlfv}
\ee 
        	
The current experimental sensitivity to slepton 
flavour mixing, i.e.\ to $( \Delta m^2_{\tilde L} )_{ij}$, 
in other observables such as $\ell_i \to \ell_j \bar\ell_k \ell_k$ 
and CR($\mu - e$, N), is in general 
smaller than in $\ell_i \to \ell_j
\gamma$~\cite{Arana-Catania:2013nha}.
The current 90\% C.L. upper-limits on the cLFV radiative decays 
are~\cite{Adam:2013mnn,Aubert:2009ag}
\be\label{eq:radiativedecays:bounds}
\text{BR($\mu\to e\gamma$)} < 5.7\times 10^{-13} \,,
\quad 
\text{BR($\tau\to\mu\gamma$)} < 4.4\times 10^{-8} \,, 
\quad
\text{BR($\tau\to e\gamma$)} < 3.3\times 10^{-8} \,.
\ee
Under the assumption that all off-diagonal entries 
$\left({Y^\nu}^\dagger L Y^\nu\right)_{ij}$ are of 
the same order of magnitude, 
the limit on BR($\mu\to e\gamma$) turns out to 
be the most constraining.
In view of this, and given the very recent experimental MEG bound on the 
BR($\mu \to e \gamma$)~\cite{Adam:2013mnn}, 
in the present update we mainly focus on the 
constraints arising from $\mu \to e \gamma$.  

\medskip
At high-energy colliders, slepton flavour mixing can be directly
probed through $\tilde \ell_{Li} \to \ell_j \chi^0_1$ 
decays~\cite{Hinchliffe:2000np,Hirsch:2008dy,Esteves:2009vg,Abada:2010kj,Abada:2011mg,Abada:2012re,Cannoni:2013gq}. 
At the LHC sleptons are preferably produced in 
cascade decays of the form $\tilde q_L \to \{ \chi^0_2, \chi^\pm_1 \}
q^\prime \to \tilde\ell_L \{\ell, \nu\} q^\prime$, provided that these
are kinematically allowed. Alternatively, sleptons can also be
present in the decay chains of directly produced wino-like $\chi^0$
and $\chi^\pm$,  
which then decay to $\tilde\ell_L$. If both these modes are not
viable, then direct production of slepton pairs through 
Drell-Yann s-channel $\gamma$ and $Z$ exchanges becomes the only 
possible slepton production mode\footnote{For a discussion of the
  prospects of LFV in slepton decays at a future Linear Collider
  see, for example,~\cite{Abada:2012re}.}. 

Despite the missing energy signature which is always present in 
R-parity conserving SUSY models with a neutral lightest SUSY particle (LSP), 
strategies to reconstruct sparticle masses have been 
devised~\cite{Hinchliffe:1996iu,Allanach:2000kt,Bachacou:1999zb}. 
These rely on the assumption that sparticles typically decay to ordinary particles 
through two body cascade decays, and that the invariant masses that can 
be formed by combining the momenta of the so-produced SM particles give rise to 
structures with edges (whose end-points 
are simple functions of sparticle masses). 
Assuming that the wino-like $\chi^0_2$ is heavier than the sleptons, 
the edge structure of the di-lepton invariant mass distributions ($m_{\ell \ell}$) 
is sensitive to slepton masses due 
to the decay $\chi^0_2 \to \tilde\ell \ell \to \ell \ell \chi^0_1$. 
An interesting effect of a high-scale SUSY seesaw is the appearance of a 
third edge in the di-lepton invariant mass distribution, due to an
intermediate slepton of a different flavour 
(i.e., $\tilde\ell_j \to \ell_i \chi^0_1$), a consequence of slepton flavour
mixing (see the detailed analysis of~\cite{Abada:2010kj}).

\subsection{Flavoured slepton mass differences} \label{sec:sms}

Here we focus on the mass differences between sleptons of different
generations (especially the first two, which are dominated by either 
the left- or right-handed slepton component).
In the absence of LFV, flavoured or inter-generational slepton mass 
differences arise from 
both $Y^l$ and in $A^l$ (with $A^l = A^l_0 Y^l$ at the GUT scale);
due to the smallness of the electron and muon Yukawa couplings 
($Y^l_{(11,22)}$), the mass differences between the 
first two generations is in general well below the $\mathcal{O}(0.1\%)$
level~\cite{Abada:2010kj}, even in the case of large $\tan\beta$.

Through RG-induced effects involving $Y^\nu$ (see the previous
subsections), the seesaw introduces additional contributions 
to slepton mass differences. 
As is clear from \eqref{eq:LLogDeltaM2}, 
these appear in the form of flavour diagonal and non-diagonal contributions to the 
slepton soft masses. Moreover, and even in the absence of
flavour-violating effects (i.e., $i=j$), their effects are manifest in
an enhancement of the fractional splittings between  $m_{\tilde e_L}$ 
and  $m_{\tilde \mu_L}$ (no significant effect in the right-handed
slepton sector, as LFV in the SUSY seesaw is mostly a left-handed phenomenon), 
which are defined as  
\begin{equation}\label{eq:sleptonMS}
\SMS\, =\, \frac{\left|m_{\tilde e_L} - m_{\tilde \mu_L}  \right|}{<
  m_{\tilde e_L}, m_{\tilde \mu_L}>}\,.
\end{equation}
Although in the presence of non-negligible flavour violation the
slepton eigenstates correspond to a mixture of the three flavours,
we will assume here that the states identified by $\tilde \ell_L$ are
dominated by the corresponding flavour component. 

Previous studies~\cite{Abada:2010kj} (before recent 2013 LHC and MEG results) 
had suggested that splittings as large as $\sim 10\%$ 
could indeed be obtained for sleptons lighter than $1~\TeV$. As mentioned 
before, we now proceed to re-evaluate these claims, in view of recent MEG bounds and 
LHC search results.

\section{Numerical results and discussion} \label{sec:results}

In our numerical analysis we assume a
normal hierarchy for the light neutrino spectrum, with 
non-vanishing $m_{\nu_1}$ (which we set $\approx \mathcal{O}(10^{-5}~\eV)$). 
The squared neutrino mass
differences, as well as the neutrino mixing angles (for a standard
parameterisation of the $\UMNS$), are taken in the intervals 
favoured by current best fits~\cite{GonzalezGarcia:2012sz}.
In our analysis we will assume vanishing CP phases (Dirac and
Majorana).

We compute the SUSY spectrum and couplings using the public code
\textsc{SPheno-3.2.2}~\cite{spheno}, extended by additional routines 
to fit the high-scale neutrino Yukawa couplings as to yield the observed 
oscillation data.  
We require that the lightest Higgs state, $h$, be compatible with
recent LHC data on a SM-like scalar boson~\cite{LHC.higgs}. 
In addition to having 
$m_h$ in the range $[123\text{ GeV},128\text{ GeV}]$, 
we further require that its couplings are not excluded 
at 95\% C.L. by current data 
using \textsc{HiggsBounds-4.0.0}~\cite{Bechtle:2013gu}. 
Concerning the sparticle spectrum, we have imposed the following
(conservative) bounds: all LEP bounds~\cite{Beringer:1900zz} were enforced; 
the gluino and first two generation squarks are required to be heavier than the
upper-limits of~\cite{ATLAS-CONF-2013-047} (derived in the limit of 
$m_{\chi^0_1} = 0$); bounds on  the 3rd generation squark
masses~\cite{LHC.3rdgeneration.squarks},
$m_{\chi^0_2}$, $m_{\chi^\pm_1}$~\cite{CMS-PAS-SUS-12-022,ATLAS-CONF-2013-028} 
and $m_{\tilde\ell}$~\cite{ATLAS-CONF-2013-049} are also
imposed (again in simplified models with $m_{\chi^0_1} = 0$). 
If the LSP 
is charged (solutions which are disfavoured in our phenomenological analysis), 
we nevertheless require its mass to be above the most constraining lower-limit 
derived from searches for heavy stable charged particles at the 
LHC~\cite{Chatrchyan:2013oca}. 
Finally, the dark-matter relic density is calculated with  
\textsc{micrOMEGAs-3.0.24}~\cite{Belanger:2013oya}.

In our analysis we first begin by considering minimal supergravity (mSUGRA) 
inspired universality conditions, and afterwards
study the impact of relaxing these universality conditions. 
This serves different purposes. First, to identify  
the regions of the parameter space offering the most promising
prospects for observation of  
slepton mass differences at forthcoming LHC
runs which, besides providing a  
SM-like Higgs, are compatible with the most recent low
energy cLFV and sparticle bounds. 
Secondly, our aim is to investigate whether the synergy 
of such mass differences (measurements or upper bounds) 
and the future results on low energy cLFV 
could still suggest some hints on the heavy 
neutrino spectrum (now rendered easier as all 
light neutrino mixing angles have been measured). 

\subsection{mSUGRA-inspired universality} \label{sec:cmssm}
	
We first consider an mSUGRA inspired framework, requiring that 
at $M_\GUT$ the SUSY soft-breaking parameters are  
universal as in cMSSM, and imposing the following relations 
on the additional seesaw soft-breaking terms: 
			\bea
			&& \left( m^2_{\tilde \nu_R} \right)_{ij} = \delta_{ij} m^2_0 \,, ~~
			\left( A^\nu \right)_{ij} = A_0 \left( Y^\nu \right)_{ij} \,,
			\eea
where $m_0$ and $A_0$ are the universal scalar soft-breaking
mass and trilinear couplings of the cMSSM. 
We begin our analysis by revisiting the 
$m_0$--$M_{1/2}$ plane, and evaluate the joint impact of the recent 
MEG bound on BR($\mu\to e\gamma$) and of the LHC negative SUSY searches. This
further allows to determine the experimentally viable regions 
offering the best prospects concerning the study of slepton mass splittings.
	
\begin{figure}[h!t]
	  \begin{center}
	    \begin{tabular}{cc}
		\includegraphics[width=80mm]{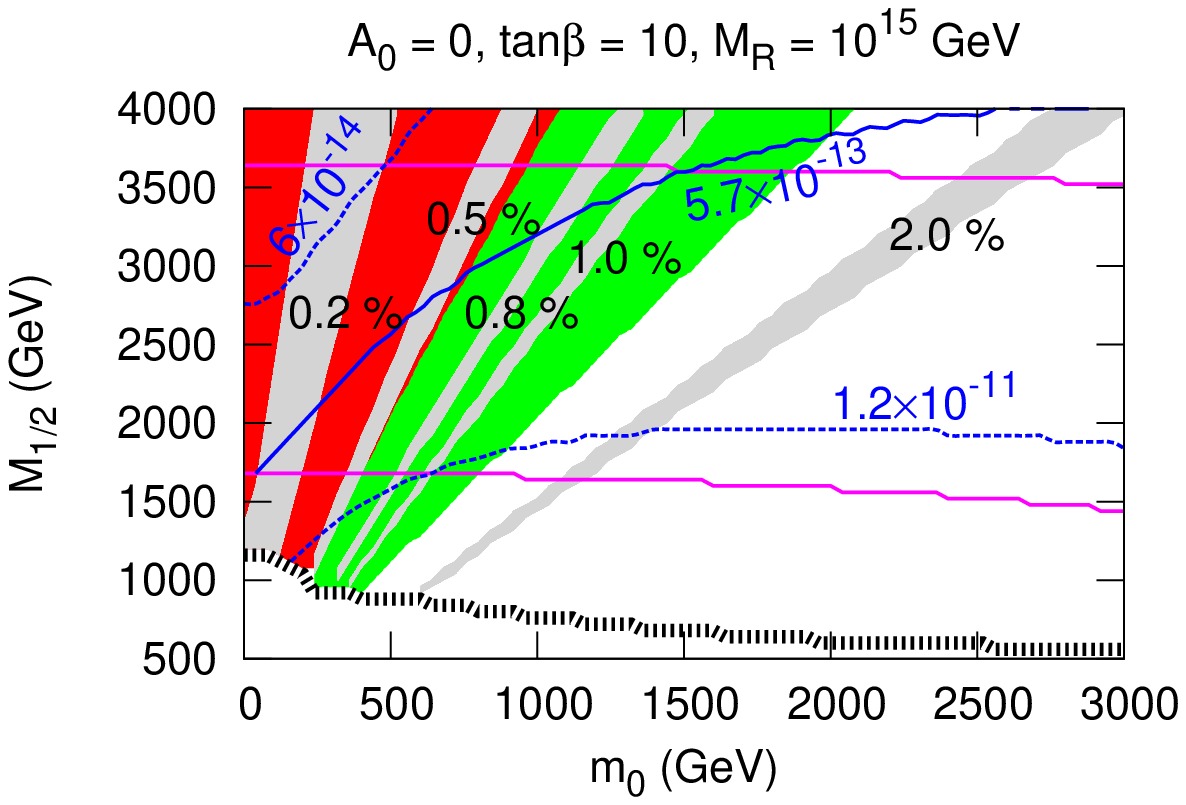} &
		\includegraphics[width=80mm]{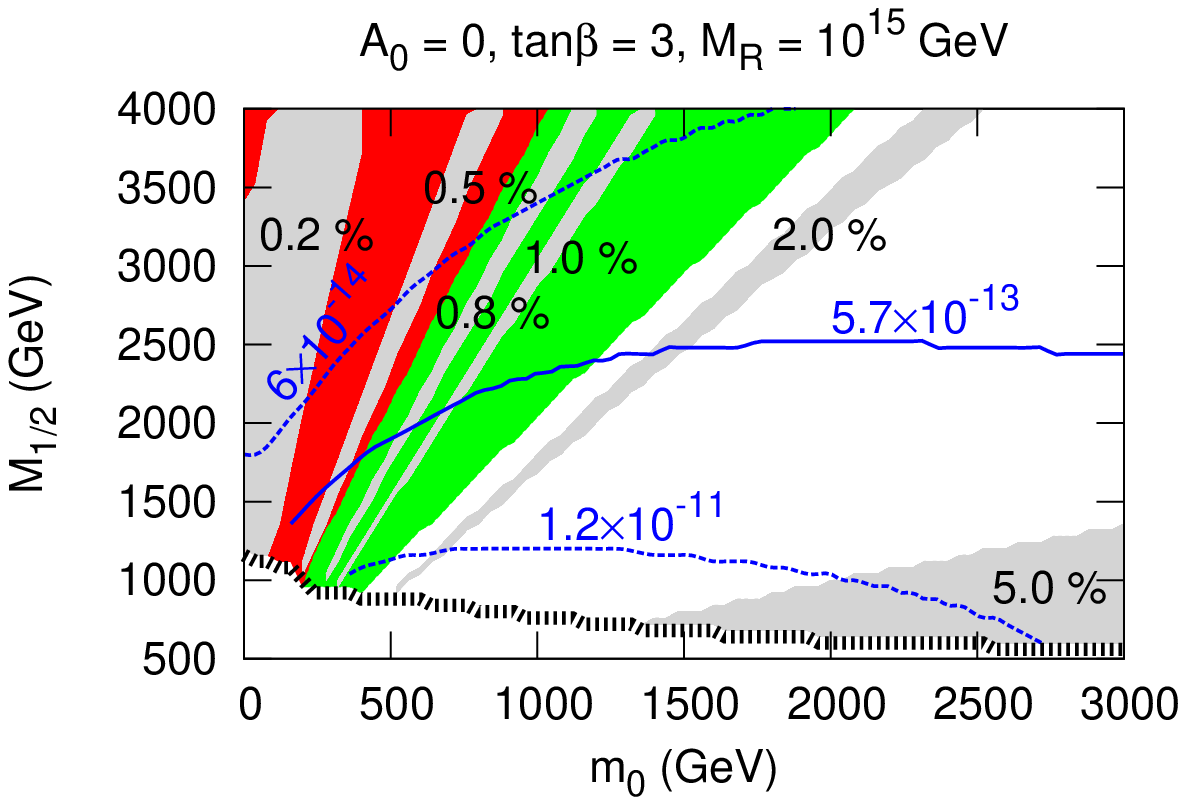} \\
		\includegraphics[width=80mm]{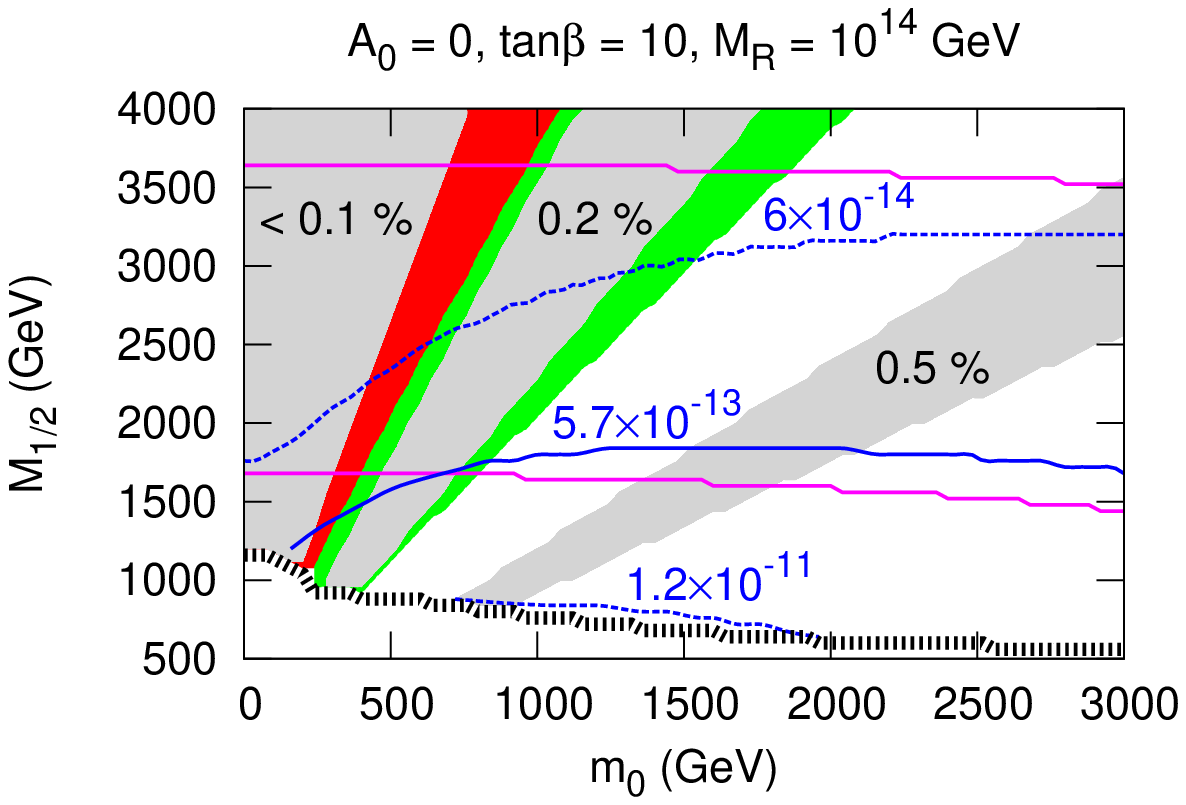} &
		\includegraphics[width=80mm]{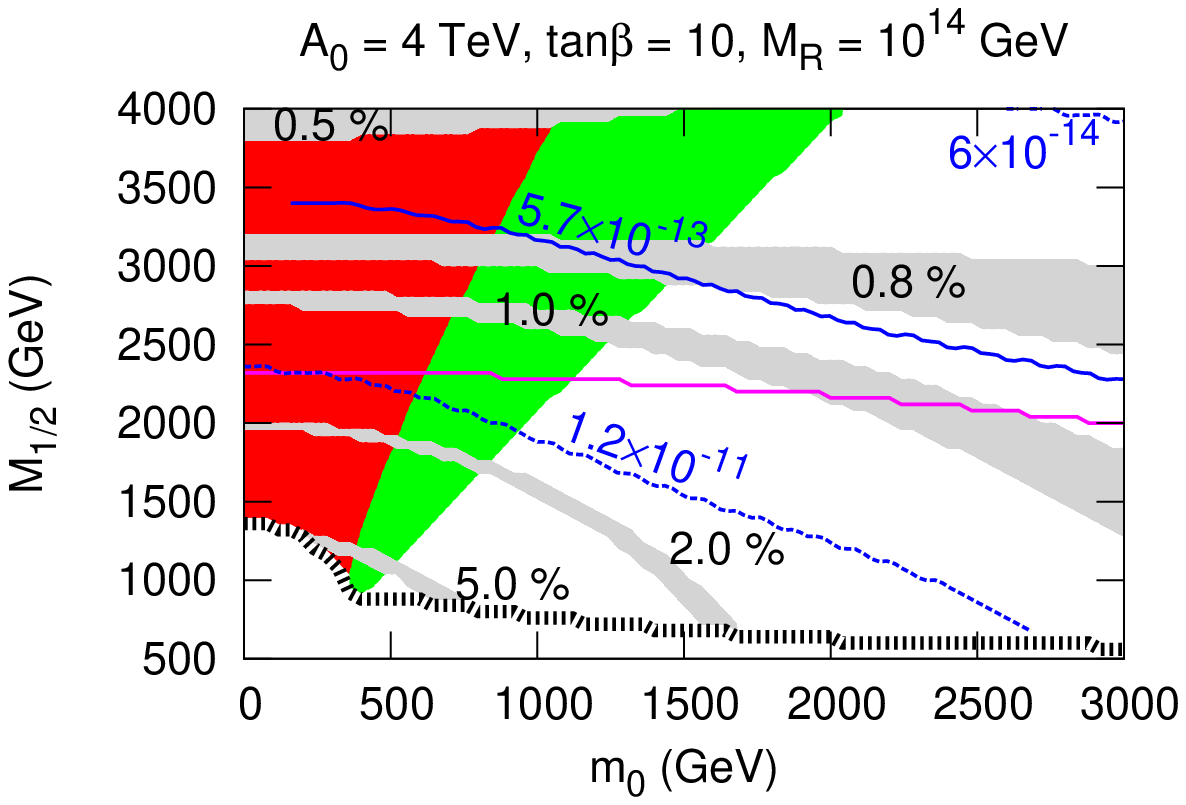}
	    \end{tabular}
	\caption{$m_0$--$M_{1/2}$ plane for different choices of
          $A_0$, $\tan \beta$ and seesaw scale ($M_R$), with different grey 
	regions corresponding to distinct values of $\SMSt$ as 
	indicated. 
	We have taken a degenerate RH neutrino spectrum 
	and set $R = \One$. 
	The region below the thick (black) dashed curve does not pass the 
	cuts applied on sparticle masses, while the red 
	regions are excluded due to the presence of a charged LSP. 
	Green regions correspond to 
	$m_{\chi^0_2,\chi^\pm_1} > \langle m_{\tilde e_L},m_{\tilde \mu_L} \rangle + 10~\GeV$. 
	Different blue curves denote BR($\mu\to e\gamma$) isolines,
	while the two solid pink lines enclose the region where $m_h
        \in [123\text{ GeV},128\text{ GeV}]$.
	}\label{fig0}
	  \end{center}
	\end{figure}

	In \figref{fig0} we present the $m_0$--$M_{1/2}$ plane\footnote{The recent analysis
          of~\cite{Camargo-Molina:2013sta} suggests that charge and
          colour breaking minima constraints on the cMSSM parameter
          space, in particular in the regime of large $|A_0|$, can be 
          more severe than previously thought. A detailed
          study of whether or not some of points here considered are
          associated to an unstable desired EWSB minimum lies beyond the 
	scope of our work. However, we expect that the large
        majority of points compatible 
	with all bounds (including flavour) falls outside the unstability regions 
	identified in~\cite{Camargo-Molina:2013sta}, where scenarios of large $|A_0|$ 
	are typically associated to very large $M_{1/2}$ and/or even larger
        $m_0$.} for different choices of $A_0$ 
	and $\tan \beta$, and for distinct seesaw
	scales. We set the $R$-matrix to $\One$ (see Eq.~(\ref{eq:CasasIbarra})),
        thus working in a conservative scenario where all
        flavour violation in the lepton sector arises from the $U^\text{MNS}$. 
	In each panel, the blue lines denote BR($\mu \to e \gamma$) isolines: the former MEGA bound,
	$1.2\times 10^{-11}$~\cite{Brooks:1999pu} (lower dashed line), 
        the MEG current bound $5.7\times 10^{-13}$~\cite{Adam:2013mnn} 
	(solid line), as well as MEG's expected future sensitivity~\cite{Baldini:2013ke}, 
	$6\times 10^{-14}$ (upper dashed line). In addition, 
	the region delimited by a thick dashed line is excluded by collider 
        bounds~\cite{ATLAS-CONF-2013-047,LHC.3rdgeneration.squarks,CMS-PAS-SUS-12-022,ATLAS-CONF-2013-028,ATLAS-CONF-2013-049,Chatrchyan:2013oca}, while the two solid 
	pink lines correspond to having $m_h \in [123\text{ GeV},128\text{ GeV}]$, in agreement
	with LHC data~\cite{LHC.higgs}. 
	\figref{fig0} clearly manifests the effect of the new BR($\mu\to
	e\gamma$) upper-limit, which in all cases amounts to 
	dramatically reducing the previously allowed cMSSM SUSY-seesaw space. 
	By lowering the seesaw scale (i.e., taking smaller values of $M_R$,
	and hence smaller $Y^\nu$),
	larger regions of the $m_0$--$M_{1/2}$ plane can survive. However, and
	as we proceed to discuss, this has a direct impact on the prospects for 
	sizeable flavoured slepton mass splittings, as both observables stem
	from a unique source of LFV - the neutrino Yukawa couplings.
	
	As can be seen in the top left panel of \figref{fig0}, the largest fractional mass
	splittings ($\SMSt$) correspond to regions with large $m_0$.
	This can be understood from the fact that (flavour non-universal) 
	RG-driven contributions to the soft masses of left-handed (LH) sleptons 
	are proportional to $m^2_{\tilde L}= m^2_{\tilde
	  \nu_R}= m^2_{H_2}=m^2_0$ at $M_\text{GUT}$
	(cf. \eqref{eq:LLogDeltaM2}). The predominant effect of a 
	larger $M_{1/2}$ translates in the increase of  
	the slepton soft-masses at the SUSY scale 
	from radiative corrections involving EW gauginos.
	For the case of a comparatively large seesaw scale ($M_R \sim
	10^{15}$ GeV), in association with large Yukawa couplings, 
	$Y^\nu \sim \mathcal{O}(1)$, one could expect
	$\SMSt \approx 2\%$; however, these regions are associated to 
	BR($\mu\to e\gamma$) already excluded. Complying with all bounds
	(accelerator - including Higgs searches -, and low-energy), 
	and further requiring a neutral LSP\footnote{In our analysis,
          and other than requiring that the LSP be neutral (typically
          the lightest neutralino), we do not impose dark matter
          constraints on the parameter space. For completeness, we
          notice that in general the relic density ($\rDM $) is always 
          $\rDM > 0.13$~\cite{Belanger:2013oya}, as points complying
          with recent bounds~\cite{Hinshaw:2012aka,Planck:2013kta} lie
          below the direct search exclusion line. One can nevertheless
          consider non-standard cosmological models, where a deviation
        from standard Big-Bang cosmology allows to reduce the relic
        density~\cite{Gelmini}, or a very small amount of R-parity violation that
        would render the LSP unstable.},  
	reduces the $m_0$--$M_{1/2}$ plane to a small triangular region, corresponding to
	$m_0 \sim 1$ TeV, $M_{1/2}\sim 3.5$ TeV, where at most one can expect 
	$\SMSt \sim \mathcal{O}(1\%)$, typically for $m_{\tilde \ell_L} \approx 2.5$ TeV. 
	
	As mentioned before, lowering the seesaw scale reduces the amount of
	RG-induced cLFV. As manifest from the comparison of the left
	panels of \figref{fig0}, the new bound on BR($\mu \to e \gamma$) can now be 
	accommodated in larger regions of the $m_0$--$M_{1/2}$ plane, but 
	slepton mass splittings also diminish, and one has 
	$\SMSt \lesssim 0.5\%$. Although this will be addressed in
        more detail in the following section, considering $R \neq
        \One$ would mostly lead to a displacement of the $\Delta
        m_{\tilde \ell}$ isosurfaces to larger values of $m_0$,
        accompanied by distortions of the  BR($\mu \to e
        \gamma$) isolines; a hierarchical RH spectrum (or fixed values of 
        $M_{R_3}$) would in turn lead to 
        a slight reduction of the associated BR($\mu\to e\gamma$). 

	The size (and global shape) of the different regions in the 
	$m_0$--$M_{1/2}$ plane also reflects the remaining mSUGRA parameters,
	$A_0$ and $\tan \beta$. The two lower panels of \figref{fig0} reflect
	the impact of varying the trilinear couplings. For 
	$| A_0 | \gg m_0$, the $A^2_0 $ contribution
	outweighs that of $m^2_0$ to the flavour non-universal radiative 
	corrections (see~\eqref{eq:LLogDeltaM2}), so that in this case
	the mass splittings are approximately constant along the $m_0$
	direction. In the regime of large $| A_0 |$, one finds that 
	the largest splittings compatible with flavour bounds are 
	$\SMSt \sim 0.8\%$, for sleptons heavier than several TeV.
	The effect of varying - in particular, lowering - $\tan
	\beta$ can be evaluated from the comparison of the two upper panels. 
        Regimes of larger $\tan \beta$ lead to larger SUSY contributions to 
        BR($\mu\to e\gamma$)~\cite{Hisano:1995nq,Hisano:1995cp}, and compatibility with current bounds
        strongly constrains the size  of the Yukawa couplings, thus
        reducing the maximal value of the slepton mass differences.
	Setting $\tan\beta = 3$ implies that in a strict cMSSM framework 
	$h$ is too light to be the SM-like Higgs. However, minimally relaxing  
	the universal conditions 
	(in particular concerning the third generation squark masses) to 
	accommodate  $m_h \sim 125$ GeV, 
	without affecting considerably the 
	observables being displayed, allows to infer
        that the low $m_0$--$M_{1/2}$ regions that were 
	excluded in the $\tan\beta=10$ case by the upper-limit 
	on $\mu\to e\gamma$ are now viable. 
	In addition, they exhibit a small enhancement of the mass 
	splittings due to an increase in the 
	strength of $Y^\nu$ ($\propto 1/v_2$, cf. \eqref{eq:CasasIbarra}).
	(Relaxed scenarios, which accommodate $m_h \sim 125$ GeV with light EW gauginos 
	and sleptons, will be explored in \secref{sec:bcmssm}.)

	To summarise the crucial point of the first part of the analysis, 
	\figref{fig0} clearly reveals how the prospects for probing the cMSSM
	type-I seesaw have evolved in view of the recent experimental
	breakthroughs. While in a first analysis\footnote{In a
          previous exploratory study~\cite{Abada:2010kj}, a regime of very small
          $\theta_{13}$ (prior to its experimental
          measurement) had been considered. For larger $\theta_{13}$, the largest slepton
          mass splittings compatible with the same set of flavour bounds 
	  are smaller than those derived for small $\theta_{13}$. In
          fact, in the case 
          $R \approx \One$ and for $M_{R_3} \gg M_{R_{1,2}}$, 
          smaller values of $\theta_{13}$ allow for a larger overall contribution to $\SMSt$ (which 
	can also proceed from $\tilde\tau - \tilde\mu$ mixing, less
        experimentally constrained).} (pre-LHC)~\cite{Abada:2010kj} one could 
	have $\SMSt \sim 5\%$ for $m_{\tilde \ell}\sim 500~\GeV$,  one is now confronted to a 
	very different situation: at most one can expect $\SMSt \sim 1\%$, and only in a
	somewhat fine-tuned region of the $m_0$--$M_{1/2}$ plane, always in
	association with considerably heavy sleptons ($\sim 2.5$~TeV).
	Whether or not the LHC will be able to reconstruct such tiny mass
	differences lies beyond the scope of the present analysis.

\begin{figure}[h!t]
	  \begin{center}
	    \begin{tabular}{cc}
		\includegraphics[width=80mm]{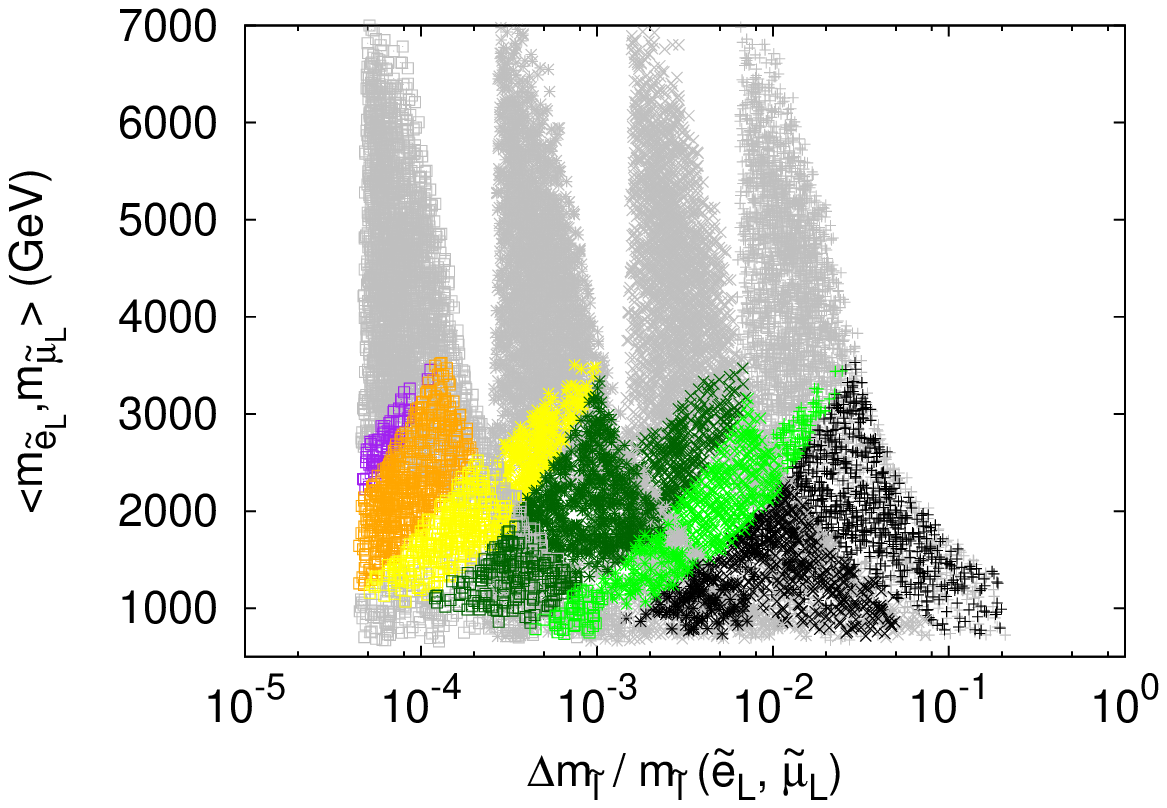} &
		\includegraphics[width=80mm]{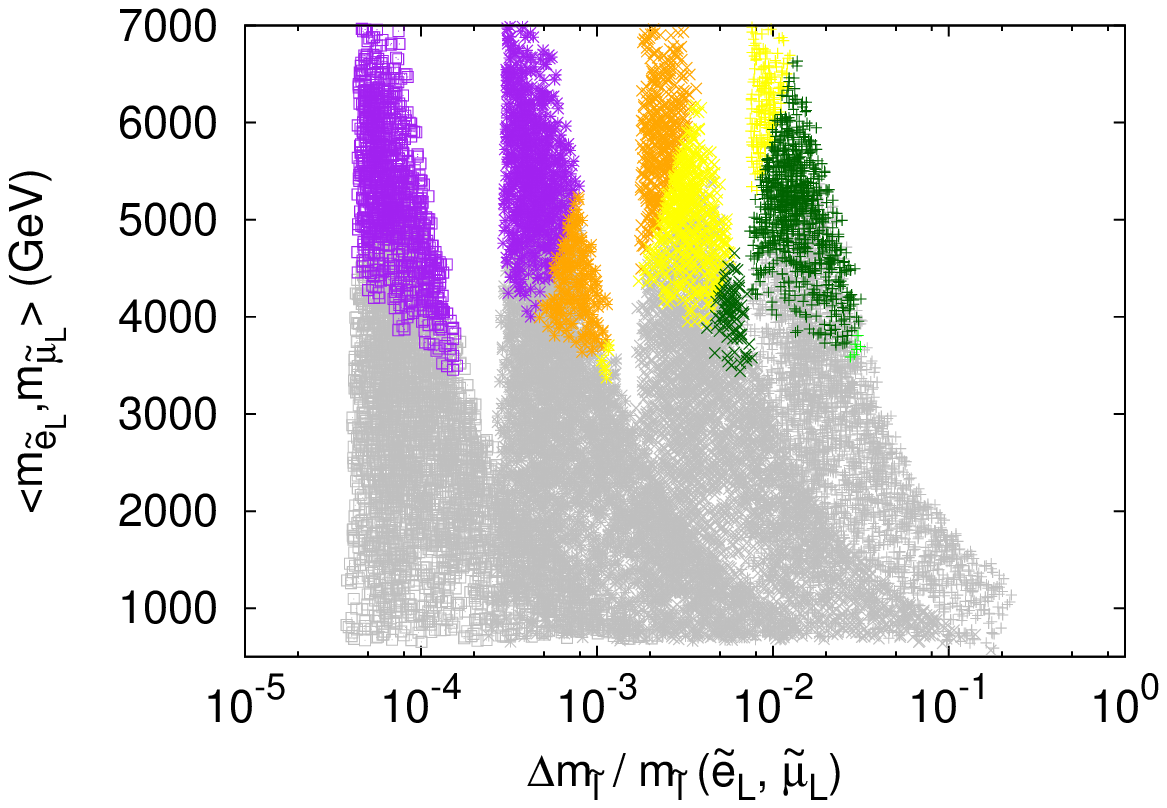} \\
	    \end{tabular}
	\caption{Slepton mass splittings versus the average slepton mass for the first 
	two generations of mostly LH sleptons, in the cMSSM type-I seesaw. On the left 
        we set $\tan \beta=10$, while on the right $\tan \beta=3$.   
        We have taken $R = \One$ and a hierarchical RH neutrino spectrum with 
	$M_{R_1} = 10^{10}$ GeV, $M_{R_2} = 10^{11}$ GeV and  
	$M_{R_3} = \{10^{12}, 10^{13}, 10^{14}, 10^{15}\}$ GeV
        (corresponding to the four regions along the $\SMSt$ axis). 
	Grey points have $m_h$ outside the preferred interval;
	purple, orange, yellow, dark-green, light-green and blac¶k regions 
        correspond to BR($\mu\to e\gamma$) in the ranges $< 10^{-17}$, 
	$[10^{-17},10^{-16}]$, $[10^{-16},10^{-15}]$, $[10^{-15},6\times 10^{-14}]$,  
	$[6\times 10^{-14},5.7\times 10^{-13}]$ and $> 5.7 \times
        10^{-13}$, respectively.}\label{fig1}
	  \end{center}
	\end{figure}

	\medskip
	As mentioned in Section~\ref{sec:cLFVobservables}, at the 
	LHC sleptons are preferably produced in 
	$\tilde q_L \to \{ \chi^0_2, \chi^\pm_1 \} \to \tilde\ell_L$
	decays (if kinematically allowed). In all panels of 
	\figref{fig0}, green surfaces correspond to regions where 
	$\tilde \ell_L$ can be produced from $\chi^0_2$ decays,
        accompanied by the emission
	of a hard lepton. We will subsequently explore these regions in
	greater detail. In particular, and instead of considering a degenerate
	RH neutrino spectrum, we now consider a 
	hierarchical one: we fix  $M_{R_1} = 10^{10}~\GeV$, $M_{R_2} = 10^{11}~\GeV$
	with  $M_{R_3} \in \{10^{12}, 10^{13}, 10^{14}, 10^{15}\}$ GeV. We
        again set $R = \One$.
	For each point we then perform a random scan in $m_0$, $M_{1/2}$ and
	$A_0$, which are chosen from the following wide ranges 
	\be
	m_0 \in [0,3] ~\TeV \,,~ M_{1/2} \in [0,10] ~\TeV \,,~ A_0 \in [-4,4] ~\TeV \,.
	\ee
	The low-energy spectrum is subject to all the 
	aforementioned cuts on sparticle (and lightest Higgs) masses.  
	One further requires that the spectrum obeys 
	$m_{\chi^0_2} > \langle m_{\tilde e_L}, m_{\tilde\mu_L} \rangle 
	+ 10~\GeV$, and that $\chi^0_1$ is the LSP.
	The results are collected in \figref{fig1}, where we display the 
	slepton mass splittings versus the average slepton mass for the first 
	two generations. 
	
	Each of the panels of \figref{fig1} comprises four
	``boomerang-shaped'' regions, corresponding to the different
	choices of $M_{R_3}$ (increasing from left to right).
	Within each individual region, the upwards (and left-most) part
	corresponds to regimes of small $| A_0 |$ (compared to $m_0$ and
	$M_{1/2}$), while the right-most extremity is associated with larger 
	$| A_0 |$.
	As it is clear from the left panel (where $\tan\beta = 10$), any splittings 
	$\gtrsim \mathcal{O}(1\%)$, compatible with current lepton flavour
	bounds, are associated to heavy sleptons, $m_{\tilde \ell} \sim
	2~\TeV$. In agreement to what could be expected from the 
	discussion\footnote{We notice that a degenerate RH neutrino
	  spectrum was considered in the analysis of \figref{fig0}. 
          For fixed values of $M_{R_3}$, a
	  hierarchical RH spectrum leads to a slight reduction of the
	  associated BR($\mu\to e\gamma$) -- by around a factor 2 --, hence
	  rendering viable part of the 1\% band in the bottom-right panel of
	  \figref{fig0}. } of \figref{fig0}, 
	these large splittings can either occur for 
	$M_{R_3} \approx 10^{15}~\GeV$ in the small $| A_0 |$ regime or 
	$M_{R_3} \approx 10^{14}~\GeV$ for large $| A_0 |$; we find it worth
	emphasising that these sizeable splittings, which have an associated 
	BR($\mu\to e\gamma$) in the range 
	of MEG's expected future sensitivity (light-green points) have a
	spectrum compatible with $m_h \sim 125~\GeV$.
	It is nevertheless possible to have $\SMSt \gtrsim 
	\mathcal{O}(0.1\%)$, compatible with current flavour bounds
	and within MEG reach, for lighter sleptons.  
        
	For comparison, on the right panel of \figref{fig1} 
	we display an analogous study for $\tan\beta = 3$. 
	As previously discussed, in such a regime, 
	complying with $m_h$ bounds requires very large values of $m_0$ and
	$M_{1/2}$ (in a strictly constrained MSSM framework), thus leading to 
	extremely heavy sleptons (and gauginos), $m_{\tilde \ell_L}\gtrsim
	3.5$ TeV. In turn, this precludes the possibility of observing
        cLFV transitions at MEG (essentially for any value of the mass splittings).

\medskip

Figure~\ref{fig1} clearly suggests that scenarios
with sizeable $\SMSt$, in association to a 
slepton spectrum sufficiently light to be abundantly produced
at the LHC, and with viable BR($\mu\to e\gamma$) are
excluded, since in a constrained framework as the cMSSM, 
such regimes are not compatible with $m_h$. 
In the following, we consider the impact of
relaxing the strict universality of the SUSY
soft-breaking terms regarding slepton mass splittings.

\subsection{Beyond mSUGRA-inspired universal conditions} \label{sec:bcmssm}
\begin{figure}[h!t]
\begin{center}
\begin{tabular}{cc}
\includegraphics[width=80mm]{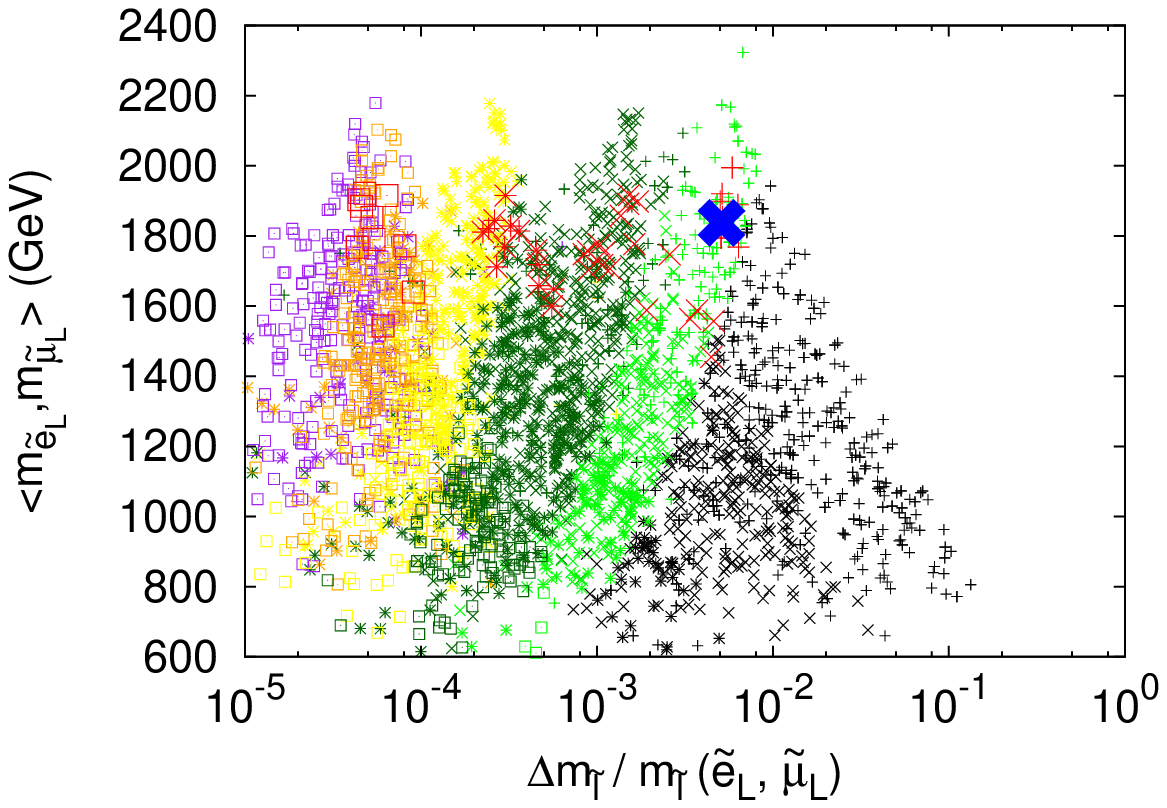} &
\includegraphics[width=80mm]{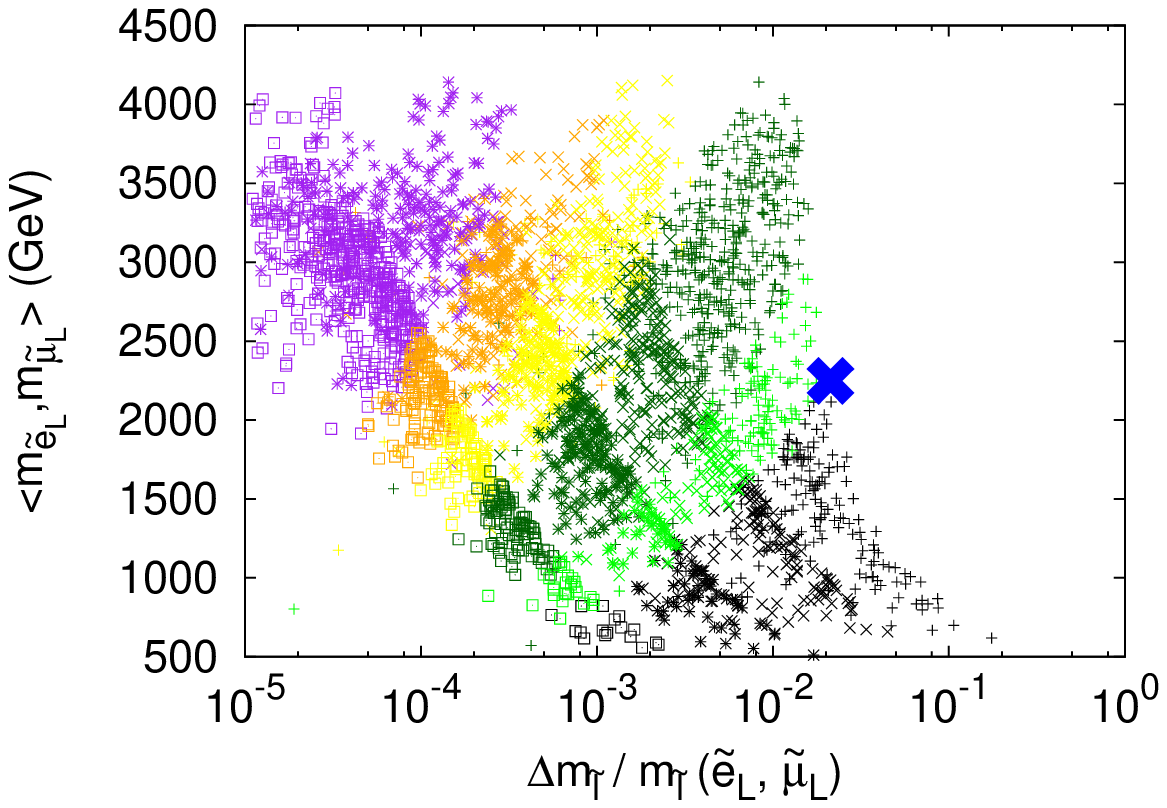}
\end{tabular}
\caption{Mass splittings versus average mass for the first 
two generations of mostly LH sleptons, for the scenario of 
\eqref{eq:cMSSM.7}.  
The underlying scan is described 
in the text, with $\tan\beta = 10$ ($3$) in the left (right) panel. We have 
taken $R = \One$, and a hierarchical RH neutrino 
spectrum with $M_{R_1} = 10^{10}$ GeV, $M_{R_2} = 10^{11}$ GeV and  
$M_{R_3} = \{10^{12}, 10^{13}, 10^{14}, 10^{15}\}$ GeV. Colour code as in
\figref{fig1}; in addition, we denote in red the points exhibiting a
DM relic density $\rDM < 0.13$. Superimposed blue ``crosses'' 
correspond to the sample points of Table~\ref{tb:bench}.}\label{fig2}
\end{center}
\end{figure}

We now consider a modified SUSY seesaw scheme in which
one breaks strict universality for squark, slepton and Higgs
soft-breaking terms at the GUT scale (but still preserving flavour 
universality). Moreover, soft breaking gluino and EW gaugino masses
are also taken to be independent. This results in the following
relations at $M_\GUT$, yielding 7 free parameters 
in addition to $\tan \beta$ and $\text{sign} (\mu)$:
\be
M_{1/2} \Rightarrow 
\left\{ \begin{array}{l}
M_1 = M_2 = M_{1/2}^{^\text{EW}} \,,  \\
M_3 = M_{1/2}^{3} \,,
\end{array} \right. 
\quad
m_0 \Rightarrow 
\left\{ \begin{array}{l}
{m}_0^{\tilde L} = {m}_0^{\tilde e} = {m}_0^{\tilde \nu_R} = m^{\tilde
  \ell}_0 \,, \vspace*{1mm}\\ \vspace*{1mm}
{m}_0^{\tilde Q} = {m}_0^{\tilde u} = {m}_0^{\tilde d} = m_0^{\tilde q} \,, \\
m_0^{H_1} = m_0^{H_2} = m_0^H \,, \\
\end{array} \right. 
\quad
A_0 \Rightarrow \left\{ \begin{array}{l}
A_0^l = A_0^\nu = A_0^\ell\,,\vspace*{1mm} \\
A_0^u = A_0^d = A_0^q \,.
\end{array} \right. 
\label{eq:cMSSM.7}
\ee
For each of the two considered regimes for $\tan \beta$, we focus on
the following specific choices for $M_{1/2}^{3}, \, m_0^{\tilde q}$ and $A_0^q$,
which lead to
$m_h \sim 125$ GeV (enhanced by radiative corrections involving heavy 
stops and/or large stop mixing), 
\bea
&\tan\beta = 10:  	& M_{1/2}^{3} = 1.1~\TeV \,, 
~ m_0^{\tilde q} = 1.5~\TeV \,, ~ A_0^q = -4~\TeV \,; \nonumber
\label{eq:colouredBOUNDARY10} \\
&\tan\beta = 3:		& M_{1/2}^{3} = 4.7~\TeV \,, 
~ m_0^{\tilde q} = 4.5~\TeV \,, ~ A_0^q = -15~\TeV \,;
\eea
and, similar to what was done for \figref{fig1}, we conduct a random
scan of the remaining parameters, which were varied in the following 
chosen intervals:
\be
M_{1/2}^{^\text{EW}} \in [0,5] ~\TeV \,, 
~ m_0^{\tilde\ell} \in [0,3] ~\TeV \,, 
~ m_0^H \in [0,3] ~\TeV \,, 
~ A_0^\ell \in [-5,5] ~\TeV \,.
\ee
The results are collected in \figref{fig2}, the panels corresponding
to $\tan \beta=10$ (left) and $\tan \beta=3$ (right). All points displayed are
in agreement with LHC bounds on sparticle masses and on a SM-like
Higgs mass. Moreover, the spectrum always fulfils 
$m_{\chi^0_2} > \langle m_{\tilde e_L}, m_{\tilde\mu_L} \rangle +
10~\GeV$, with a $\chi^0_1$ LSP. 

As it is manifest from \figref{fig2}, 
the most interesting consequence of relaxing the strict cMSSM universality
conditions concerns 
the possibility of having considerably lighter sleptons in association
with sizeable mass splittings and still compatible with flavour bounds. 
This is particularly striking in the case of $\tan\beta = 3$ (right
panel), where one can verify that sleptons as light as 
$\sim 800~\GeV$ ($1.6~\TeV$) can be associated to $\mathcal{O}(0.1\%)$
($\mathcal{O}(1\%)$) splittings, in agreement with all imposed
constraints. 
For $\tan\beta = 10$ (displayed on the left panel), the impact of
deviating from a strict cMSSM framework is somewhat less pronounced:
one again finds that $\mathcal{O}(0.1\%)$ slepton mass splittings are 
attainable for $m_{\tilde \ell} \gtrsim 0.9~\TeV$.
Larger values of $\SMSt$ remain difficult to obtain with the boundary 
conditions of \eqref{eq:colouredBOUNDARY10}. 

A very interesting feature of this relaxed framework, especially for
the $\tan\beta = 10$ case, is that the spectrum now allows for an
efficient LSP density depletion (via $\chi_1^0 - \tilde t_1$
co-annihilation). This is highlighted by the red points in the left
panel of \figref{fig2}.

Concerning the prospects of a potential future observation of a 
$\mu \to e \gamma$ decay at MEG, in the case of $\tan\beta = 10$, 
and for sleptons lighter than $1.2~\TeV$, 
$\SMSt \gtrsim \mathcal{O}(0.1\%)$ are associated to 
BR($\mu\to e\gamma$) within MEG's expected future sensitivity.
For $\tan\beta = 3$, splittings $\gtrsim \mathcal{O}(1\%)$ ($\mathcal{O}(0.1\%)$), 
with sleptons lighter than $\sim 2.7~\TeV$ ($1.3~\TeV$),  would also yield
BR($\mu\to e\gamma$) values within MEG reach. 

As compared to \figref{fig1}, where one could still identify four
independent ``boomerang'' shapes (especially for heavy sleptons), here 
only four small crests can be distinguished (again for the heavier
slepton regimes). Due to having uncorrelated
$m_0^{\tilde q}$ and $m_0^{H}$ at the GUT scale, and having taken heavy
stops (with large $A_0^{u}$), 
$m^2_{H_2}$ can now run to negative values above the seesaw 
scale, thus potentially cancelling the contribution of
$m^2_{\tilde\nu_R}$, $m^2_{\tilde L}$ and $|A_0^\nu|^2$   
to the flavour violating RG-induced effects (cf. \eqref{eq:LLogDeltaM2}). 
Potential cancellations within the
flavour-violating soft-breaking terms were pointed out
in~\cite{Calibbi:2012gr} for the case of non-universal Higgs 
masses (where possible negative values of $m^2_{H_2}$ at the GUT scale were 
considered). 
Finally, we notice that points in the left (right) panel have an
average squark mass for the first two LH generations 
$m_{\tilde q_L} \sim 3 \,(9)$~TeV. Although this
formally means that all points displayed in \figref{fig2} would have  
the $\tilde q_L \to \chi^0_2 \to \tilde\ell_L$ cascade open, 
in the general case the most likely slepton production mode remains
via direct gaugino 
production (due to the very heavy squark spectrum).

\medskip
In what follows we evaluate to which extent 
the correlations between low- and high-energy cLFV
observables can still provide potential probes of a type-I SUSY
seesaw, illustrating the results for two individual cases, singled out
from the panels of \figref{fig2}, 
where they have been depicted using blue ``crosses''. 

Point A (left panel - $\tan \beta=10$) exhibits the largest splittings 
for a light slepton spectrum, and has $\rDM$ within 
the $1\sigma$ interval of Planck~\cite{Planck:2013kta}. Point B
(right panel - $\tan \beta=3$) also corresponds to a choice
illustrating the largest splittings still compatible with 
current flavour bounds.  
In Table~\ref{tb:bench} we display the 
SUSY soft breaking terms (in addition to the fixed input parameters of 
\eqref{eq:colouredBOUNDARY10}), and 
a sample of the SUSY spectrum - corresponding to $M_{R_3} =10^{15}$
GeV. Strictly for illustrative purposes, and using \textsc{Prospino-2.1}~\cite{prospino}, 
we also provide an estimation of the LH slepton production
cross-sections for  $\sqrt s = 14~\TeV$, from the decay of 
directly produced neutral and charged winos
($\chi^0_2, \,\chi^\pm_1$), as well as from the decay chains of 
directly produced squarks and gluinos. 
(Notice that Point A is a concrete example of a case where slepton production 
via squark production is more favourable than via direct gaugino production.)
\begin{table}[h!t]
\centering
\renewcommand{\arraystretch}{1.4}
\hspace*{-3mm}
\begin{tabular}{|c|c|c|}
\hline
& A & B \\\hline
$M^{^\text{EW}}_{1/2}$ & 2362 & 2821 \\\hline
$m^{\tilde\ell}_0$ & 973 & 1365 \\\hline
$m^H_0$ & 173 & 2808 \\\hline
$A^\ell_0$ & 3750 & 1651 \\\hline
\end{tabular}
\renewcommand{\arraystretch}{1.0}
\hspace*{3mm}
\renewcommand{\arraystretch}{1.4}
\begin{tabular}{|c|c|c|}
\hline
& A & B \\\hline
$\langle m_{\tilde e_L}, m_{\tilde\mu_L} \rangle$ & 1835 & 2240 \\\hline
$\frac{\Delta m_{\tilde\ell}}{m_{\tilde\ell\phantom{y}}}
(\tilde e_L,\tilde\mu_L)$ & 0.5\% & 2.1\% \\\hline
$m_{\chi^0_2,\chi^\pm_{1\phantom{y}}}$ & 1936 & 2349 \\\hline
$m_{\chi^0_{1\phantom{y}}}$ & 1043 & 1264 \\\hline
$m_{\tilde t_1}$ & 1084 & 4825 \\\hline
$\langle m_{\tilde q_L} \rangle$ & 2916 & 9059 \\\hline
$m_h$ &125.4 & 125.3 \\\hline
\end{tabular}
\renewcommand{\arraystretch}{1.0}
\vspace*{3mm}
\renewcommand{\arraystretch}{1.4}
\begin{tabular}{|l|c|c|}
\hline
& A & B \\\hline
$\sigma_{\tilde \ell \phantom{y}}^{\text{prod } (\chi^0_2, \chi^\pm_1)}$ & 
$6.1 \times 10^{-4} $ & 
$7.6 \times 10^{-4}$ 
\\\hline
$\sigma_{\tilde \ell \phantom{y}}^{\text{prod } (\chi^0_2 \text{ only})}$ & 
$2.1 \times 10^{-4}$ & 
$2.6 \times 10^{-4}$ 
\\\hline
$\sigma_{\tilde \ell \phantom{y}}^{\text{prod } (\tilde q_L \text{; via } \chi^0_2,
\chi^\pm_1)}$ & 
$1.9 \times 10^{-2}$ 
& --
\\\hline
$\sigma_{\tilde \ell \phantom{y}}^{\text{prod } (\tilde q_L \text{; via
}\chi^0_2 \text{ only})}$ &  
$6.4 \times 10^{-3}$ 
& --
\\\hline
\end{tabular}
\renewcommand{\arraystretch}{1.0}
\caption{Points A and B: SUSY soft breaking input parameters, 
sample of the SUSY spectrum and slepton mass splittings (in the 
case of $M_{R_3} = 10^{15}$ GeV), and LH slepton production
cross-sections for $\sqrt s = 14~\TeV$ (masses in 
GeV and $\sigma$ in fb).}\label{tb:bench}
\end{table}

Although the production cross-sections for these points suggest that
observation of a significant number of events might be challenging, it
is important to stress that a minor modification of the spectra
(in particular, breaking the universality of the third generation of squarks)
could easily lead to scenarios where as much as 300 events can be indeed achieved, for an
integrated luminosity around 3000 fb$^{-1}$. 

\begin{figure}[h!t]
\begin{center}
\begin{tabular}{cc}
\includegraphics[width=80mm]{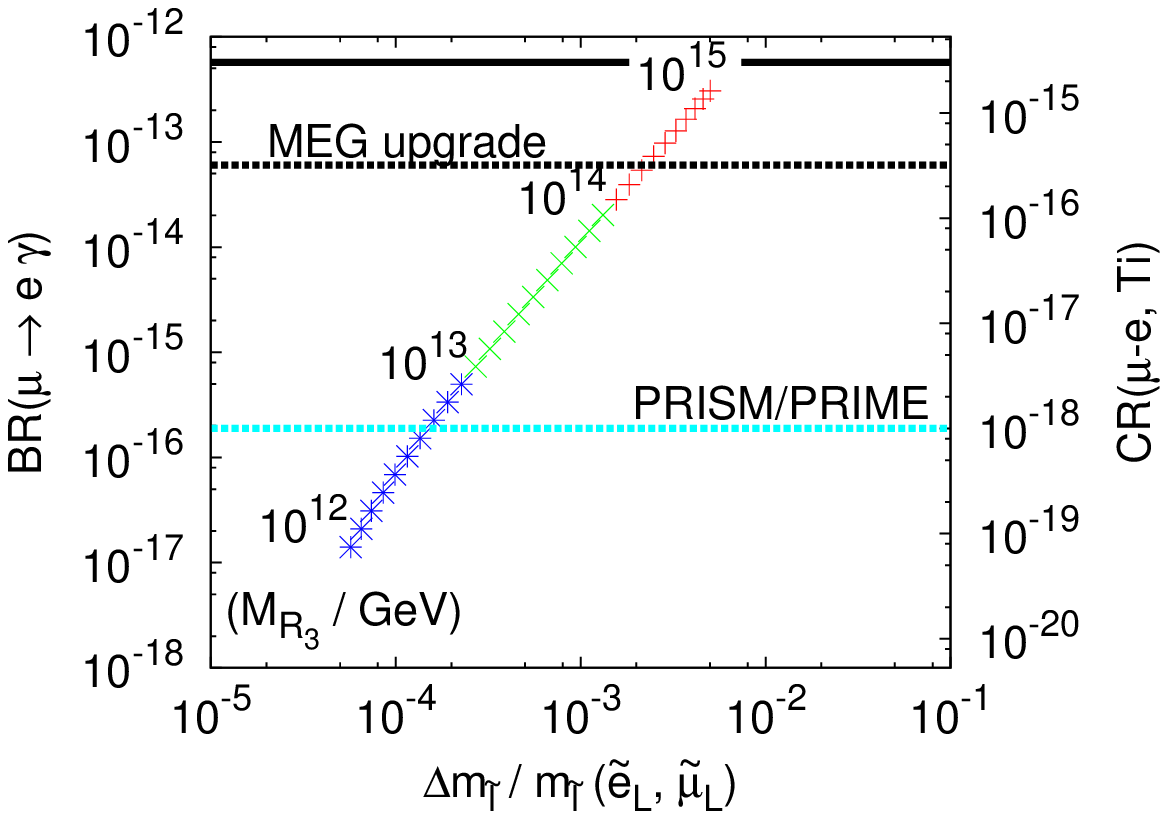} &
\includegraphics[width=80mm]{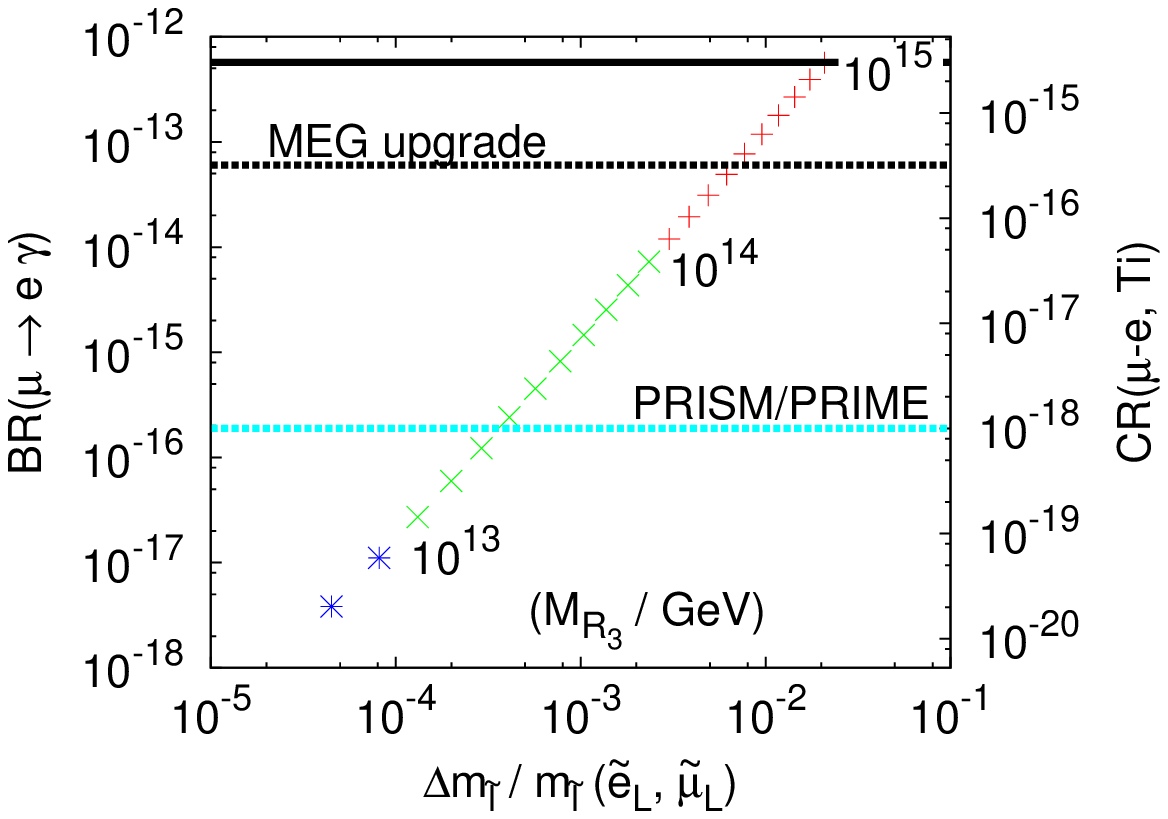} \\
\includegraphics[width=80mm]{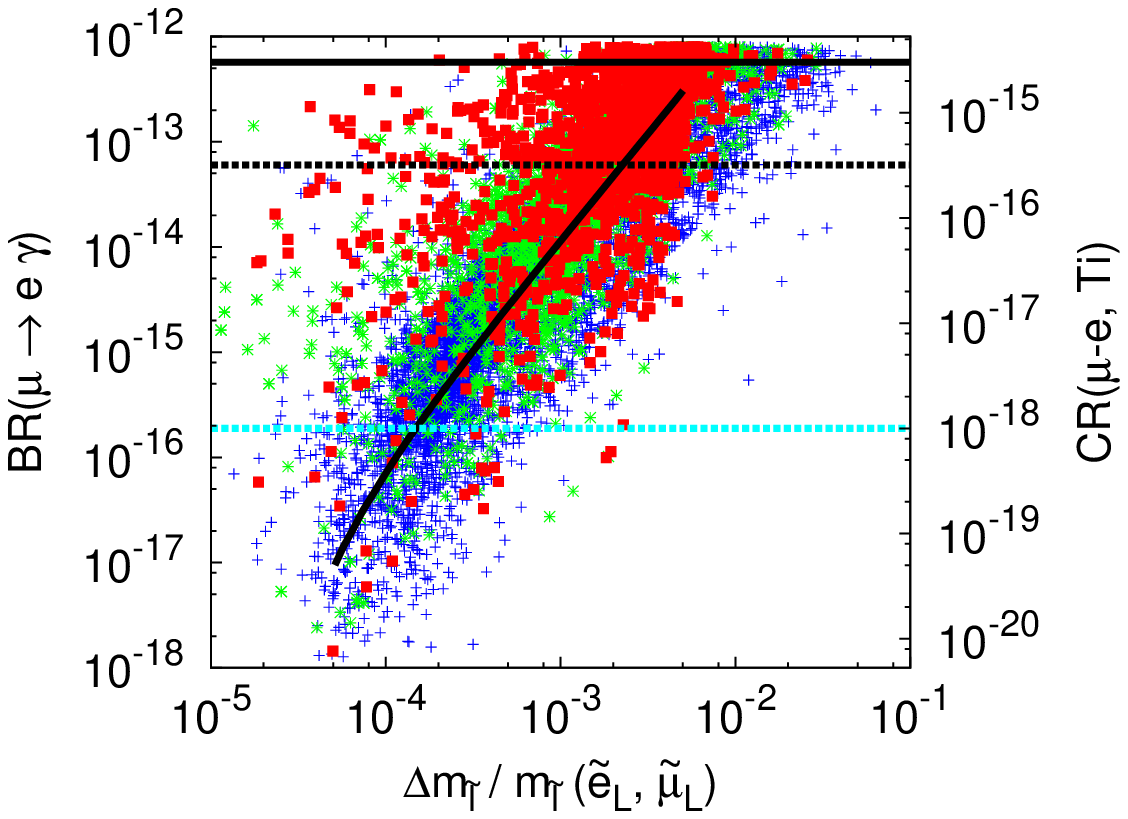} &
\includegraphics[width=80mm]{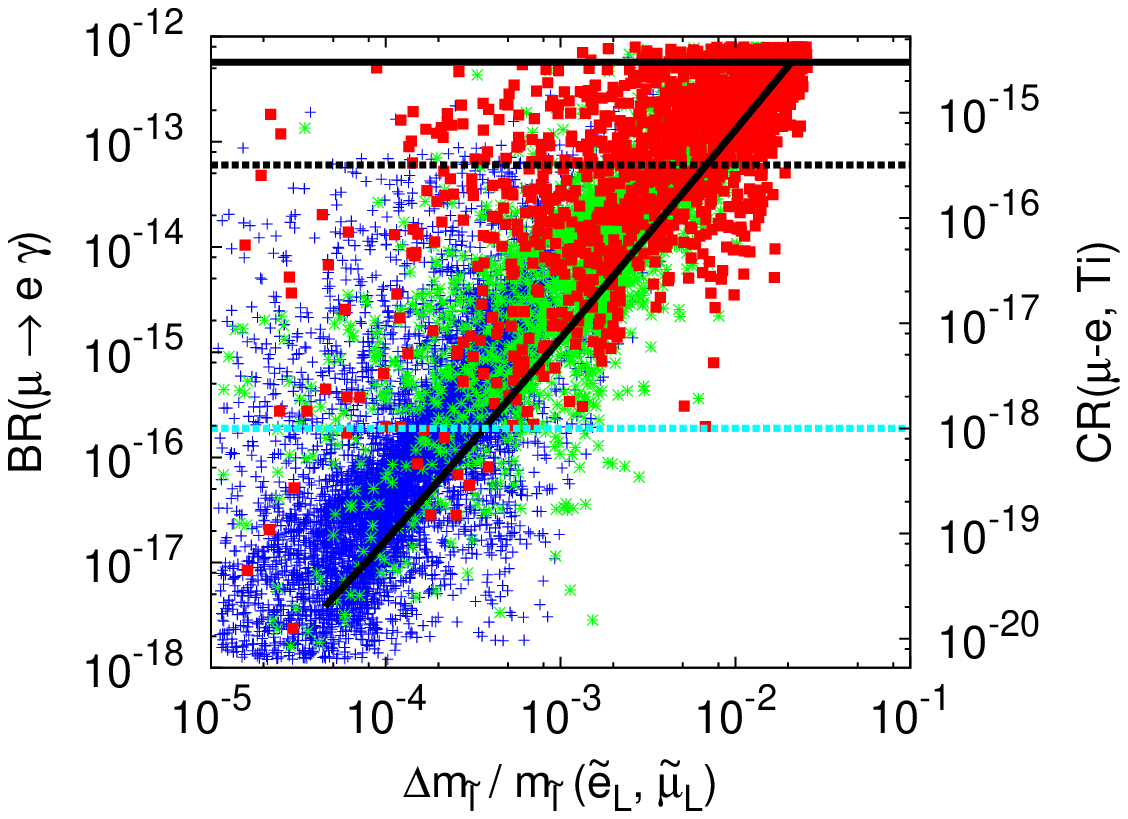} \\
\end{tabular}
\caption{Mass splittings versus BR($\mu\to e\gamma$) for points A  
(left) and B (right) - see Table~\ref{tb:bench}, displaying complementary
information on CR($\mu - e$, Ti) on the secondary y-axis. Full
(dashed) horizontal lines denote current bounds (future sensitivities).
We have taken a hierarchical RH neutrino spectrum with $M_{R_1} =
10^{10}$ GeV, $M_{R_2} = 10^{11}$ GeV. The colour scheme denotes
different intervals of $M_{R_3}$: in the upper panels it is varied 
in the range $[10^{12},10^{15}]$ GeV, while in the lower panels we 
considered $10^{13,14,15}$ GeV (blue, green and red 
points, respectively). On the upper panels, we set 
$R = \One$, while in the lower ones the $R$-matrix was randomly varied  
in the intervals $|\theta_i| \lesssim \pi$ and 
$-\pi \lesssim \text{arg}(\theta_i) \lesssim \pi$.
The slanted full black lines on the lower panels correspond to the
limit $R = \One$ depicted in the upper panels.}\label{fig3}
\end{center}
\end{figure}
In \figref{fig3} we display BR($\mu\to e\gamma$) as a function of
$\SMSt$ for points A (left panel) and B (right panel), considering
different regimes of $M_{R_3}$ (for fixed values of $M_{R_{1,2}}$). On
the secondary y-axis we present the corresponding value of 
CR($\mu - e$, Ti), estimated assuming the hypothesis of
$\gamma$-penguin domination, valid for the scenarios here
considered~(see, for
example\cite{Arganda:2007jw,Arana-Catania:2013nha}), and 
which predicts 
CR($\mu - e$, Ti) $\approx 5 \times 10^{-3}$ BR($\mu\to e\gamma$)~\cite{Hisano:2001qz}.
A horizontal full (dashed) line corresponds to MEG's current bound
(expected future sensitivity), while a cyan dashed line corresponds 
to PRISM/PRIME proposed sensitivity\footnote{Other proposals for
high-sensitivity $\mu-e$ conversion searches (Aluminium nuclei) include
Mu2e~\cite{Gaponenko:2012fx} and COMET~\cite{Kuno:2013mha}.}, 
CR($\mu - e$, Ti) $\sim 10^{-18}$~\cite{Barlow:2011zza}. 
On the upper two plots, we have again taken 
the conservative - yet simple - limit of $R = \One$. On the lower plots of 
\figref{fig3}, we consider the more general case where there are
additional mixings involving the RH neutrinos, conducting a broad scan
over the complex $R$-matrix angles, $\theta_i$. This allows a
global overview of the contributions to the different cLFV
observables (albeit for fixed SUSY points).

The information contained in the different panels of \figref{fig3} 
provides a comprehensive summary of the discussion we have conducted
so far. Taking into account the additional degrees of freedom in the RH neutrino
sector (encoded in the $R$-matrix complex angles) allows to have 
mass splittings as large as 5\%, still in agreement with low-energy
flavour bounds, and even for a comparatively low seesaw scale,
$\mathcal{O}(10^{13}~\GeV)$, as made clear from the lower panels. 
Strong deviations from the simplistic $R = \One$ 
case may also point in the direction of particular flavour 
models, as recently explored in~\cite{Cannoni:2013gq}.

\medskip
From the interplay of the different observables, and depending
on the outcome of the distinct low- and high-energy experiments, many 
conclusions can be drawn with respect to the viability of a type-I
SUSY seesaw as the underlying mechanism.

Under the hypothesis that the slepton mass scale will have been determined
(and the sparticle spectrum interpreted in terms of a high-scale SUSY model),
a joint study of the low- and high-energy cLFV
observables can offer important insight into the seesaw dynamics. 
Let us first consider the simple $R = \One$ case:
for scenario A, any measurement of $\SMSt
\gtrsim 0.1\%$ would entail BR($\mu\to e\gamma$) within MEG reach (and
vice-versa),
further suggesting a seesaw scale of the order of $10^{14}~\GeV$. 
For reconstructed SUSY models of lower $\tan \beta$, the result in
the upper right panel would allow for similar conclusions: 
splittings of $\mathcal{O}(1\%)$ 
($\mathcal{O}(0.1\%)$) should be accompanied by the observation of $\mu\to
e\gamma$ decay at MEG ($\mu- e$ conversion at PRISM/PRIME), hinting towards 
a seesaw scale of the order of $5 \times 10^{14}~\GeV$ ($10^{14}~\GeV$).
Conversely, the isolated manifestation of either low- or high-scale
cLFV, e.g. $\SMSt \gtrsim \mathcal{O}(5\%)$ without any $\mu\to e\gamma$ or 
$\mu -e $ signal, would strongly suggest that sources of LFV, other
than - or in addition to - the SUSY seesaw are present.

Analogous, but stronger conclusions can be drawn from the inspection
of the scans corresponding to $R \neq \One$ case: although the 
clear dependence on the seesaw scale (present for $R = \One$) 
becomes diluted due to the additional
contributions from the mixings involving RH neutrinos, 
the correlation between the
observables still allows to indirectly test the SUSY seesaw.
Again, any measurement $\SMSt \gtrsim \mathcal{O}(1\%)$ must 
be accompanied by observation of $\mu\to e\gamma$ decay at MEG
so to substantiate the SUSY seesaw hypothesis; on the other hand  
$\SMSt \gtrsim \mathcal{O}(1\%)$ without any $\mu\to e\gamma$ or 
$\mu -e $ signal would strongly disfavour the underlying hypothesis.

\section{Conclusions} \label{sec:concl}

In this study we have revisited the impact of a type-I SUSY seesaw
concerning LFV following the recent MEG bound on BR($\mu \to e
\gamma$), LHC data (discovery of a Higgs-like boson and negative SUSY
searches), and the measurement of $\theta_{13}$, updating the results
obtained in~\cite{Abada:2010kj}. 
The aim of our work was to discuss whether current cLFV results and
SUSY searches still render viable the observation of slepton mass
differences at the LHC, and if the interplay of the latter
observables with low-energy cLFV bounds could still shed some light
on the high-energy seesaw parameters.
Our analysis was based in the hypothesis that all flavour violation in
the lepton sector is due to the neutrino Yukawa couplings; 
we thus embed the type-I seesaw in constrained and semi-constrained  
SUSY breaking scenarios. 

Due to the new BR($\mu\to e\gamma$) bound, 
in association with the measured ``large''
value of $\theta_{13}$, we find that in general slepton mass
splittings tend to be very small, unless the slepton spectra is
considerably heavy. This implies that for the type-I SUSY seesaw 
the observation
of cLFV at high-energies will be clearly much more challenging than 
the low-energy, high-intensity studies.

Regarding the embedding of the type-I seesaw into constrained SUSY
models such as the cMSSM, we have verified that  
recent LHC data (in particular the measurement of $m_h$) 
precludes the possibility of
simultaneously having BR($\mu\to e\gamma$) within MEG reach and 
sizeable slepton mass differences associated with a 
slepton spectrum sufficiently light to be produced. 

On the other hand, relaxing the strict universality of SUSY
soft-breaking terms allows to circumvent some of the strongest LHC
bounds, especially on $m_h$, and opens the door to non-negligible 
slepton mass splittings (for a comparatively light
slepton spectrum), with associated $\mu\to e\gamma$ rates (as well as
CR($\mu - e $)) within experimental reach. 
Although dependent on the SUSY regime (e.g., on $\tan \beta$), one can
still find $\SMSt \sim 0.1\% - 1\%$, for $m_{\tilde \ell}$ ranging
from 800~GeV to 1.6~TeV. 

We have studied 
in detail the impact of the different seesaw parameters
for representative points in SUSY space. The results
of this comprehensive analysis were presented in \figref{fig3}. 
As we have shown, one can have
mass splittings as large as 5\%, still in agreement with low-energy
flavour bounds, even for a comparatively low seesaw scale, 
$\mathcal{O}(10^{13}~\GeV)$. In these scenarios, one can still
use the correlation of high- and low-energy cLFV observables to probe
the SUSY seesaw, and we provided some illustrative examples of this
interplay. 

In summary, our analysis shows that in the case of semi-constrained
(flavour universal) SUSY models, the reconstruction of the slepton
mass scale and slepton mass splittings, in synergy with the
measurement of low-energy cLFV observables, still 
remains a potential probe to test
(strengthen or disfavour) the high-scale type-I SUSY seesaw.

\section*{Acknowledgements}
We are indebted to A. Abada and J. C. Rom\~ao for many
valuable exchanges and suggestions. 
The work of  A. J. R. F.  has been supported by {\it Funda\c c\~ao
  para a Ci\^encia e a 
Tecnologia} through the fellowship SFRH/BD/64666/2009. 
A. J. R. F. acknowledges the financial support from
the EU Network grant UNILHC PITN-GA-2009-237920 and from {\it
Funda\c{c}\~ao para a Ci\^encia e a Tecnologia} grants CFTP-FCT UNIT
777, CERN/FP/83503/2008 and PTDC/FIS/102120/2008. 
We also acknowledge partial support from the
European Union FP7  ITN INVISIBLES (Marie Curie Actions, PITN-GA-2011-289442).


\begin{thebibliography}{99}

\bibitem{Borzumati:1986qx}
  F.~Borzumati and A.~Masiero,
  Phys.\ Rev.\ Lett.\  {\bf 57} (1986) 961.

\bibitem{seesaw:I}
  P.~Minkowski,
  Phys.\ Lett.\ B {\bf 67} (1977) 421;
  %
  M.~Gell-Mann, P.~Ramond and R.~Slansky, in {\it Complex Spinors and
  Unified Theories} eds. P.~Van.~Nieuwenhuizen and D.~Z.~Freedman,
  {\it Supergravity} (North-Holland, Amsterdam, 1979), 
  p.315 [Print-80-0576 (CERN)];
  %
  T.~Yanagida, in {\it Proceedings of the Workshop on the Unified Theory
  and the Baryon Number in the Universe}, eds. O.~Sawada and
  A.~Sugamoto (KEK, Tsukuba, 1979), p.95;
  %
  S.~L.~Glashow, in {\it Quarks and Leptons}, eds. M.~L\'evy {\it et
  al.} (Plenum Press, New York, 1980), p.687;
  %
  R.~N.~Mohapatra and G.~Senjanovi\'c,
  Phys.\ Rev.\ Lett.\  {\bf 44} (1980) 912.

\bibitem{Hisano:1995cp}
  J.~Hisano, T.~Moroi, K.~Tobe and M.~Yamaguchi,
  Phys.\ Rev.\ D {\bf 53} (1996) 2442
  [hep-ph/9510309].

\bibitem{Hisano:1995nq}
  J.~Hisano, T.~Moroi, K.~Tobe, M.~Yamaguchi and T.~Yanagida,
  Phys.\ Lett.\ B {\bf 357} (1995) 579
  [hep-ph/9501407].

\bibitem{Hisano:1998fj}
  J.~Hisano and D.~Nomura,
  Phys.\ Rev.\  D {\bf 59} (1999) 116005 
  [arXiv:hep-ph/9810479].

\bibitem{Buchmuller:1999gd}
  W.~Buchmuller, D.~Delepine and F.~Vissani,
  Phys.\ Lett.\  B {\bf 459} (1999) 171
  [arXiv:hep-ph/9904219].
 
\bibitem{Kuno:1999jp}
  Y.~Kuno and Y.~Okada,
  Rev.\ Mod.\ Phys.\  {\bf 73} (2001) 151
  [arXiv:hep-ph/9909265].

\bibitem{Ellis:1999uq}
  J.~R.~Ellis, M.~E.~Gomez, G.~K.~Leontaris, S.~Lola and D.~V.~Nanopoulos,
  Eur.\ Phys.\ J.\ C {\bf 14} (2000) 319
  [hep-ph/9911459].

\bibitem{Hisano:2001qz}
  J.~Hisano and K.~Tobe,
  Phys.\ Lett.\ B {\bf 510} (2001) 197
  [hep-ph/0102315].

\bibitem{Casas:2001sr}
  J.~A.~Casas and A.~Ibarra,
  Nucl.\ Phys.\ B {\bf 618} (2001) 171
  [hep-ph/0103065].

\bibitem{Lavignac:2001vp}
  S.~Lavignac, I.~Masina and C.~A.~Savoy,
  Phys.\ Lett.\  B {\bf 520} (2001) 269
  [arXiv:hep-ph/0106245].

\bibitem{Bi:2001tb}
  X.~J.~Bi and Y.~B.~Dai,
  Phys.\ Rev.\  D {\bf 66} (2002) 076006 
  [arXiv:hep-ph/0112077].

\bibitem{Ellis:2002fe}
  J.~R.~Ellis, J.~Hisano, M.~Raidal and Y.~Shimizu,
  Phys.\ Rev.\ D {\bf 66} (2002) 115013
  [hep-ph/0206110].

\bibitem{Deppisch:2002vz}
  F. Deppisch, H. Pas, A. Redelbach, R. Ruckl and Y. Shimizu,
  Eur. Phys. J. C {\bf 28} (2003) 365
  [arXiv:hep-ph/0206122].

\bibitem{Fukuyama:2003hn}
  T.~Fukuyama, T.~Kikuchi and N.~Okada,
  Phys.\ Rev.\  D {\bf 68}  (2003) 033012
  [arXiv:hep-ph/0304190].

\bibitem{Brignole:2004ah}
  A.~Brignole and A.~Rossi,
  Nucl.\ Phys.\  B {\bf 701}  (2004) 3 
  [arXiv:hep-ph/0404211].

\bibitem{Masiero:2004js}
  A.~Masiero, S.~K.~Vempati and O.~Vives,
  New J.\ Phys.\  {\bf 6} (2004) 202
  [arXiv:hep-ph/0407325].

\bibitem{Fukuyama:2005bh}
  T.~Fukuyama, A.~Ilakovac and T.~Kikuchi,
  Eur.\ Phys.\ J.\  C {\bf 56} (2008) 125
  [arXiv:hep-ph/0506295].

\bibitem{Petcov:2005jh}
  S.~T.~Petcov, W.~Rodejohann, T.~Shindou and Y.~Takanishi,
  Nucl.\ Phys.\  B {\bf 739} (2006) 208
  [arXiv:hep-ph/0510404].

\bibitem{Arganda:2005ji}
  E.~Arganda and M.~J.~Herrero,
  Phys.\ Rev.\ D {\bf 73} (2006) 055003
  [hep-ph/0510405].

\bibitem{Deppisch:2005rv}
  F.~Deppisch, H.~Pas, A.~Redelbach and R.~Ruckl,
  Phys.\ Rev.\ D {\bf 73} (2006) 033004
  [hep-ph/0511062].

\bibitem{Yaguna:2005qn}
  C.~E.~Yaguna,
  Int.\ J.\ Mod.\ Phys.\  A {\bf 21} (2006) 1283
  [arXiv:hep-ph/0502014].

\bibitem{Calibbi:2006nq}
  L.~Calibbi, A.~Faccia, A.~Masiero and S.~K.~Vempati,
  Phys.\ Rev.\  D {\bf 74} (2006) 116002 
  [arXiv:hep-ph/0605139].

\bibitem{Antusch:2006vw}
  S.~Antusch, E.~Arganda, M.~J.~Herrero and A.~M.~Teixeira,
  JHEP {\bf 0611} (2006) 090
  [hep-ph/0607263].

\bibitem{Arganda:2007jw}
  E.~Arganda, M.~J.~Herrero and A.~M.~Teixeira,
  JHEP {\bf 0710} (2007) 104
  [arXiv:0707.2955 [hep-ph]].

\bibitem{Arganda:2008jj}
  E.~Arganda, M.~J.~Herrero and J.~Portoles,
  JHEP {\bf 0806} (2008) 079
  [arXiv:0803.2039 [hep-ph]].

\bibitem{Arkanihamed:1996au}
  N. Arkani-Hamed, H. Cheng, J. L. Feng and L. J. Hall,
  Phys. Rev. Lett. {\bf 77} (1996) 1937
  [arXiv:hep-ph/9603431].

\bibitem{Hinchliffe:2000np}
  I. Hinchliffe and F. E. Paige,
  Phys. Rev. D {\bf 63} (2001) 115006
 [arXiv:hep-ph/0010086].

\bibitem{Carvalho:2002jg}
  D. F. Carvalho, J. R. Ellis, M. E. Gomez, S. Lola and J. C. Romao,
  Phys. Lett. B {\bf 618} (2005) 162
  [arXiv:hep-ph/0206148].

\bibitem{Buckley:2006nv}
  M. R. Buckley and H. Murayama,
  Phys. Rev. Lett. {\bf 97} (2006) 231801
  [arXiv:hep-ph/0606088].

\bibitem{Hirsch:2008dy}
  M. Hirsch, J. W. F. Valle, W. Porod, J. C. Romao and A. Villanova del Moral,
  Phys. Rev. D {\bf 78} (2008) 013006
  [arXiv:0804.4072].

\bibitem{Carquin:2008gv}
  E. Carquin, J. Ellis, M. E. Gomez, S. Lola and J. Rodriguez-Quintero,
  JHEP {\bf 0905} (2009) 026
  [arXiv:0812.4243].

\bibitem{Esteves:2009vg}
  J.~N.~Esteves, J.~C.~Romao, A.~Villanova del Moral, M.~Hirsch,
  J.~W.~F.~Valle and W.~Porod, 
  JHEP {\bf 0905} (2009) 003
  [arXiv:0903.1408 [hep-ph]].

\bibitem{Buras:2009sg}
  A. J. Buras, L. Calibbi and P. Paradisi, 
  JHEP {\bf 1009} (2010) 042
  [arXiv:0912.1309].

\bibitem{Abada:2010kj}
  A.~Abada, A.~J.~R.~Figueiredo, J.~C.~Romao and A.~M.~Teixeira,
  JHEP {\bf 1010} (2010) 104
  [arXiv:1007.4833 [hep-ph]].
  
\bibitem{Abada:2011mg}
  A.~Abada, A.~J.~R.~Figueiredo, J.~C.~Romao and A.~M.~Teixeira,
  JHEP {\bf 1108} (2011) 099
  [arXiv:1104.3962 [hep-ph]].

\bibitem{Calibbi:2011dn}
  L.~Calibbi, R.~N.~Hodgkinson, J.~Jones Perez, A.~Masiero and O.~Vives,
  Eur.\ Phys.\ J.\ C {\bf 72} (2012) 1863
  [arXiv:1111.0176 [hep-ph]].

\bibitem{Galon:2011wh}
  I.~Galon and Y.~Shadmi,
  Phys.\ Rev.\ D {\bf 85} (2012) 015010
  [arXiv:1108.2220 [hep-ph]].

\bibitem{Arbelaez:2011bb}
  C.~Arbelaez, M.~Hirsch and L.~Reichert,
  JHEP {\bf 1202} (2012) 112
  [arXiv:1112.4771 [hep-ph]].

\bibitem{Abada:2012re}
  A.~Abada, A.~J.~R.~Figueiredo, J.~C.~Romao and A.~M.~Teixeira,
  JHEP {\bf 1208} (2012) 138
  [arXiv:1206.2306 [hep-ph]].

\bibitem{Cannoni:2013gq}
  M.~Cannoni, J.~Ellis, M.~E.~Gomez and S.~Lola,
  Phys.\ Rev.\ D {\bf 88} (2013) 075005
  [arXiv:1301.6002 [hep-ph]].

\bibitem{Abe:2011sj}
  K.~Abe {\it et al.}  [T2K Collaboration],
  Phys.\ Rev.\ Lett.\  {\bf 107} (2011) 041801
  [arXiv:1106.2822 [hep-ex]].

\bibitem{Abe:2011fz}
  Y.~Abe {\it et al.}  [DOUBLE-CHOOZ Collaboration],
  Phys.\ Rev.\ Lett.\  {\bf 108} (2012) 131801
  [arXiv:1112.6353 [hep-ex]].

\bibitem{Ahn:2012nd}
  J.~K.~Ahn {\it et al.}  [RENO Collaboration],
  Phys.\ Rev.\ Lett.\  {\bf 108} (2012) 191802
  [arXiv:1204.0626 [hep-ex]].

\bibitem{An:2012bu}
  F.~P.~An {\it et al.}  [Daya Bay Collaboration],
  Chin.\  Phys.\ C {\bf 37} (2013) 011001
  [arXiv:1210.6327 [hep-ex]].

\bibitem{ATLAS-CONF-2013-047}
  G.~Aad {\it et al.}  [ATLAS Collaboration],
  ``Search for squarks and gluinos with the ATLAS detector in final
  states with jets and missing transverse momentum and 20.3 fb$^{-1}$
  of $\sqrt{s}=8$ TeV proton-proton collision data,'' 
  ATLAS-CONF-2013-047.

\bibitem{LHC.3rdgeneration.squarks}
  S.~Chatrchyan {\it et al.}  [CMS Collaboration],
  ``Search for top-squark pair production in the single lepton final
  state in pp collisions at 8 TeV,''
  CMS-PAS-SUS-13-011;
%
  G.~Aad {\it et al.}  [ATLAS Collaboration],
  ``Search for direct production of the top squark in the all-hadronic
  ttbar + etmiss final state in 21 fb-1 of p-p collisions at sqrt(s)=8
  TeV with the ATLAS detector,'' 
  ATLAS-CONF-2013-024;
%
  G.~Aad {\it et al.}  [ATLAS Collaboration],
  ``Search for direct third generation squark pair production in final
  states with missing transverse momentum and two $b$-jets in
  $\sqrt{s}$ = 8 TeV $pp$ collisions with the ATLAS detector,'' 
  ATLAS-CONF-2013-053;
%
  G.~Aad {\it et al.}  [ATLAS Collaboration],
  ``Search for strongly produced superpartners in final states with
  two same sign leptons with the ATLAS detector using 21 fb-1 of
  proton-proton collisions at sqrt(s)=8 TeV,'' 
  ATLAS-CONF-2013-007.

\bibitem{CMS-PAS-SUS-12-022}
  S.~Chatrchyan {\it et al.}  [CMS Collaboration],
  ``Search for direct EWK production of SUSY particles in multilepton
  modes with 8TeV data,'' 
  CMS-PAS-SUS-12-022.

\bibitem{ATLAS-CONF-2013-028}
  G.~Aad {\it et al.}  [ATLAS Collaboration],
  ``Search for electroweak production of supersymmetric particles in
  final states with at least two hadronically decaying taus and
  missing transverse momentum with the ATLAS detector in proton-proton
  collisions at $\sqrt{s} = 8$ TeV,''  
  ATLAS-CONF-2013-028.

\bibitem{ATLAS-CONF-2013-049}
  G.~Aad {\it et al.}  [ATLAS Collaboration],
  ``Search for direct-slepton and direct-chargino production in final
  states with two opposite-sign leptons, missing transverse momentum
  and no jets in 20/fb of pp collisions at sqrt(s) = 8 TeV with the
  ATLAS detector,'' 
  ATLAS-CONF-2013-049.  

\bibitem{Chatrchyan:2013oca}
  S.~Chatrchyan {\it et al.}  [CMS Collaboration],
  JHEP {\bf 07} (2013) 122
  arXiv:1305.0491 [hep-ex].

\bibitem{LHC.higgs}
  G.~Aad {\it et al.}  [ATLAS Collaboration],
  Phys.\ Lett.\ B {\bf 716} (2012) 1
  [arXiv:1207.7214 [hep-ex]];
  G.~Aad {\it et al.}  [ATLAS Collaboration],
  ``Combined measurements of the mass and signal strength of the
  Higgs-like boson with the ATLAS detector using up to 25 fb$^{-1}$ of
  proton-proton collision data,'' 
  ATLAS-CONF-2013-014;
  G.~Aad {\it et al.}  [ATLAS Collaboration],
  ``Combined coupling measurements of the Higgs-like boson with the
  ATLAS detector using up to 25 fb$^{-1}$ of proton-proton collision
  data,'' 
  ATLAS-CONF-2013-034;
  S.~Chatrchyan {\it et al.}  [CMS Collaboration],
  JHEP {\bf 06} (2013) 081
  [arXiv:1303.4571 [hep-ex]].

\bibitem{Adam:2013mnn}
  J.~Adam {\it et al.}  [MEG Collaboration],
  Phys. Rev. Lett. {\bf 110} (2013) 201801
  arXiv:1303.0754 [hep-ex].

\bibitem{Heuer:2012gi}
  R.-D.~Heuer,
  ``News from CERN, LHC Status and Strategy for Linear Colliders,''
  arXiv:1202.5860 [physics.acc-ph].

\bibitem{Barr:1512933}
  A.~Barr, K.~Boone, A.~Canepa, M.~Crispin, M.~D'Onofrio, C.~Young,
  G.~Polesello and G.~Redlinger, 
  ``Searches for Supersymmetry at the high luminosity LHC with the
  ATLAS Detector,'' 
  ATL-PHYS-PUB-2013-002.

\bibitem{Calibbi:2012gr}
  L.~Calibbi, D.~Chowdhury, A.~Masiero, K.~M.~Patel and S.~K.~Vempati,
  JHEP {\bf 1211} (2012) 040
  [arXiv:1207.7227].

\bibitem{Grossman:1997is}
  Y.~Grossman and H.~E.~Haber,
  Phys.\ Rev.\ Lett.\  {\bf 78} (1997) 3438
  [hep-ph/9702421].

\bibitem{Raidal:2008jk}
  M.~Raidal {\it et al.},
  Eur.\ Phys.\ J.\  C {\bf 57} (2008) 13
  [arXiv:0801.1826 [hep-ph]] and references therein.

\bibitem{Arana-Catania:2013nha}
  M.~Arana-Catania, S.~Heinemeyer and M.~J.~Herrero,
  Phys.\ Rev.\ D {\bf 88} (2013) 015026
  [arXiv:1304.2783 [hep-ph]].

\bibitem{Aubert:2009ag}
  B.~Aubert {\it et al.}  [BaBar Collaboration],
  Phys.\ Rev.\ Lett.\  {\bf 104} (2010) 021802
  [arXiv:0908.2381 [hep-ex]].

\bibitem{Hinchliffe:1996iu}
  I.~Hinchliffe, F.~E.~Paige, M.~D.~Shapiro, J.~Soderqvist and W.~Yao,
  Phys.\ Rev.\ D {\bf 55} (1997) 5520
  [hep-ph/9610544].

\bibitem{Allanach:2000kt}
  B.~C.~Allanach, C.~G.~Lester, M.~A.~Parker and B.~R.~Webber,
  JHEP {\bf 0009} (2000) 004
  [hep-ph/0007009].

\bibitem{Bachacou:1999zb}
  H.~Bachacou, I.~Hinchliffe and F.~E.~Paige,
  Phys.\ Rev.\ D {\bf 62} (2000) 015009
  [hep-ph/9907518].

\bibitem{GonzalezGarcia:2012sz}
  M.~C.~Gonzalez-Garcia, M.~Maltoni, J.~Salvado and T.~Schwetz,
  JHEP {\bf 1212} (2012) 123
  [arXiv:1209.3023 [hep-ph]].

\bibitem{spheno}
  W.~Porod,
  Comput.\ Phys.\ Commun.\  {\bf 153} (2003) 275
  [hep-ph/0301101].
  W.~Porod and F.~Staub,
  Comput.\ Phys.\ Commun.\  {\bf 183} (2012) 2458
  [arXiv:1104.1573 [hep-ph]].

\bibitem{Bechtle:2013gu}
  P.~Bechtle, O.~Brein, S.~Heinemeyer, O.~Stal, T.~Stefaniak,
  G.~Weiglein and K.~Williams, 
  PoS CHARGED {\bf 2012} (2012) 024
  [arXiv:1301.2345 [hep-ph]].

\bibitem{Beringer:1900zz}
  J.~Beringer {\it et al.}  [Particle Data Group Collaboration],
  Phys.\ Rev.\ D {\bf 86} (2012) 010001.

\bibitem{Belanger:2013oya}
  G.~Belanger, F.~Boudjema, A.~Pukhov and A.~Semenov,
  ``micrOMEGAs3.1 : a program for calculating dark matter observables,''
  arXiv:1305.0237 [hep-ph].

\bibitem{Camargo-Molina:2013sta}
  J.~E.~Camargo-Molina, B.~O'Leary, W.~Porod and F.~Staub,
  ``Stability of the CMSSM against sfermion VEVs,''
  arXiv:1309.7212 [hep-ph].

\bibitem{Brooks:1999pu}
  M.~L.~Brooks {\it et al.}  [MEGA Collaboration],
  Phys.\ Rev.\ Lett.\  {\bf 83} (1999) 1521
  [hep-ex/9905013].

\bibitem{Baldini:2013ke}
  A.~M.~Baldini {\it et al.}, 
  ``MEG Upgrade Proposal,''
  arXiv:1301.7225 [physics.ins-det].

\bibitem{Planck:2013kta}
  P.~A.~R.~Ade {\it et al.}  [Planck Collaboration],
  ``Planck 2013 results. XV. CMB power spectra and likelihood,''
  arXiv:1303.5075 [astro-ph.CO].

\bibitem{Hinshaw:2012aka}
  G.~Hinshaw {\it et al.}  [WMAP Collaboration],
  Astrophys.\ J.\ Suppl.\  {\bf 208} (2013) 19
  arXiv:1212.5226 [astro-ph.CO].

\bibitem{Gelmini}
  G.~Gelmini, P.~Gondolo, A.~Soldatenko and C.~E.~Yaguna,
  Phys.\ Rev.\ D {\bf 74} (2006) 083514
  [hep-ph/0605016].
  G.~B.~Gelmini and P.~Gondolo,
  Phys.\ Rev.\ D {\bf 74} (2006) 023510
  [hep-ph/0602230].

\bibitem{prospino}
  W.~Beenakker, R.~Hopker, M.~Spira and P.~M.~Zerwas,
  Nucl.\ Phys.\ B {\bf 492} (1997) 51
  [hep-ph/9610490];
  W.~Beenakker, M.~Klasen, M.~Kramer, T.~Plehn, M.~Spira and P.~M.~Zerwas,
  Phys.\ Rev.\ Lett.\  {\bf 83} (1999) 3780
  [Erratum-ibid.\  {\bf 100} (2008) 029901]
  [hep-ph/9906298],
  Phys.\ Rev.\ Lett.\  {\bf 83} (1999) 3780
  [Erratum-ibid.\  {\bf 100} (2008) 029901]
  [hep-ph/9906298].

\bibitem{Gaponenko:2012fx}
  A.~Gaponenko [mu2e Collaboration],
  ``The Mu2e Experiment: A New High-Sensitivity Muon to Electron
  Conversion Search at Fermilab,'' 
  FERMILAB-CONF-12-490-PPD.

\bibitem{Kuno:2013mha}
  Y.~Kuno [COMET Collaboration],
  PTEP {\bf 2013} (2013) 022C01.

\bibitem{Barlow:2011zza}
  R.~J.~Barlow,
  Nucl.\ Phys.\ Proc.\ Suppl.\  {\bf 218} (2011) 44.

\end{thebibliography}
\end{document}